\documentclass[eqsecnum]{emulateapj}

\usepackage{natbib,amsmath}
\usepackage{epsfig}
\usepackage{hyperref}

\usepackage{parskip}
\shortauthors{Wani et al.}

\begin{document}

\shorttitle{X-ray studies of Blazar 1ES 1959$+$650 in June 2018--December 2020}

\title{X-ray studies of Blazar 1ES 1959+650 using Swift \& XMM--Newton satellite }

\author{Kiran A Wani\altaffilmark{1,2}, Haritma Gaur\altaffilmark{1} and M K Patil\altaffilmark{2} }
\altaffiltext{1}{Aryabhatta Research Institute of Observational Sciences (ARIES), Manora Peak, Nainital - 263 002, India;
harry.gaur31@gmail.com; kiranwani189@gmail.com}
\altaffiltext{2}{School of Physical Sciences, SRTM University, Nanded, 431 606, India}

\begin{abstract}
\noindent
High synchrotron energy peaked blazar 1ES 1959$+$650 is studied with Swift and XMM--Newton \hbox{satellite} in total 127 observations
during the period June 2018$-$December 2020. We extensively studied its flux and spectral variability on intra-day and long-term
timescales. Discrete correlation function analysis between soft and hard X-ray bands indicates soft as well as hard lags. The
results are used to constrain the magnetic field of the emitting region which is found to be 0.64$\pm$0.05 Gauss. On long-term timescales, 
distribution of fluxes shows lognormality behaviour which could be attributed to minijets-in-a-jet model or might be due to the propagation of relativistic
shocks down the jet. The spectral energy distribution
around the synchrotron peak is well described by the log parabola model. Spectral parameters like peak energy $E_{p}$,
curvature $\beta$ and the peak luminosity $L_{p}$ are derived from spectral analysis. Their correlations are studied to constrain
the acceleration processes of the emitting particles. $E_{p}$ shows strong correlation with $L_{p}$ during the high
state of the source which indicates spectral changes might be caused by the variations of the average electron energy.  
Low values of curvature parameter $\beta$ and a weak correlation between $E_{p}$ and $\beta$ indicates co-existence of 
stochastic/statistical acceleration of electrons in the emitting region. Implications of other results are also discussed.
\end{abstract}


\section{Introduction}
\label{sec:introduction}
\noindent
Blazar, a subclass of AGN, constitutes BL Lacertae objects (BL Lacs) and flat spectrum radio quasars (FSRQs). Blazars
are characterized by rapid flux variability at time scales ranging from a few minutes to years across the entire electromagnetic spectrum; high optical polarisation and the emission is predominantly non-thermal in nature which emanates from a relativistic jet streaming along or aligned very close to our line of sight \citep{Blandford1978}.

The spectral energy distribution of blazars consists of double peaked hump with the first one peaking in sub-mm to  soft
X-rays, whereas the second hump peaks at MeV to TeV energies \citep{Urry_1995}. The low energy component of the SED
is mostly due to the synchrotron emission from the relativistic electrons. However, the physical mechanisms behind the high
energy emission are not well understood and are thought to be originating from the Compton upscattering of synchrotron
photons by same population of relativistic electrons (Synchrotron self Compton, SSC i.e. \citealt{Ghisellini1989, Mastichiadis1997}) or external seed photons originating from the accretion disc, broad line region (BLR) and torus 
components of a blazar (External Compton, EC i.e. \citealt{Dermer1992, Ghisellini1998}). The other models which 
appear to be viable mechanisms to explain the X-ray through $\gamma$-ray emission are hadronic models where the high energy emission is produced by relativistic protons through proton synchrotron radiation and photo-pion production, followed by pion decay and electromagnetic cascades \cite[e.g.][and references therein]{Mannheim1992, Bottcher2013a}.

Blazars are divided into three classes based on the position of synchrotron peak in their SED \citep{Padovani_1995}. Low frequency  BL  Lac  objects  exhibit synchrotron  peak  in  IR-Optical  band,  
intermediate  frequency BL  Lac  objects  have  their  synchrotron  peak  at  optical-UV frequencies and high frequency BL Lac 
(HBL) objects show a synchrotron peak in the UV to X-ray band.

1ES 1959+650 is at a redshift of z=0.048 \citep{Perlman_1996} and is a prominent high synchrotron peaked blazar in which 
the synchrotron peak of the SED appears in the UV-X-ray band (\citealt{Krawczynski_2004, Kapanadze2016, Abdo_2010}).
In X-rays, it was first detected by the Slew Survey with the Einstein Imaging Proportional Counter (IPC) (i.e. \citealt{Elvis_1992}),
followed by BeppoSAX \citep{Beckmann_2002}, RXTE, Swift, XMM--Newton (\citealt{Tagliaferri_2003, Massaro(2008)}), by 
the Nuclear Spectroscopic Telescope Array (NuSTAR) \citep{Pandey2017} in later years.

It has shown strong flux variability in optical, X-ray and TeV energy bands \cite[i.e.][]{Krawczynski_2004, Kapanadze2016, Kapanadze2018a,Kapanadze2018b, Kaur_2017, Patel_2018, Wang_2018, Wang_2019, Magic2020}. High X-ray flaring 
activity of the source has been reported by \cite{Perlman_2005} and \cite{Krawczynski_2004} using XMM--Newton and RXTE-PCA 
observations in 2002--2003. \cite{Kapanadze2016} reported frequent X-ray flares of this source during 2005--2014 using Swift--XRT observations.

1ES 1959+650 underwent an unprecedented X-ray flaring activity during August 2015--January 2016. During this period, it varied 
by a factor of $\sim$5.7 with maximum value above 20 counts/sec, along with high flux activity in TeV energy band (i.e. \citealt{Kapanadze2016, Kaur_2017, Patel_2018}). However, in several 
multi-wavelength campaigns, orphan flares in $\gamma$-rays (which are not simultaneous with X-ray flares) have been found in June 2002 (i.e. \citealt{Krawczynski_2004}) and found that 
the orphan $\gamma$-ray flare cannot be explained with conventional one-zone SSC models. They invoked Multiple-Component SSC models; External Compton models where the variations of the external
photon intensity in the jet frame can cause orphan $\gamma$-ray flares; magnetic field aligned along jet axis and thus the observer
would not see the synchrotron flare but the electrons would scatter SSC $\gamma$-rays in our direction and thus be able to see the inverse Compton flare. Bottcher (2005) proposed
that orphan flares are difficult to reconcile with the standard leptonic SSC model and suggested that they may originate from relativistic protons interacting with an external
photon field supplied by electron synchrotron radiation reflected off a dilute reflector. \cite{Chandra_2021} modelled the SED of this blazar using a single zone 
time dependent SSC model which could explain the flares successfully. \cite{Patel_2018} studied this source during observation period June-July 2016 and explained its broadband SED using two zone SSC model where the inner zone is mainly responsible for producing the synchrotron peak and the high energy $\gamma$-ray part whereas, the second zone explains less variable optical-UV and low energy $\gamma$-ray emission.

The X-ray spectral index hardens with increasing flux level in the long-term duration ( \citealt{Patel_2018}; \citealt{Magic2020}). and also during a number of flares (i.e. \citealt{Wang_2018}). \cite{Kapanadze2016, Kapanadze2018b, Kapanadze2018a} also showed the ‘harder-when-
brighter’ trend in blazar 1ES 1959+650.
Recently, \cite{Chandra_2021} presented an extensive analysis of 1ES 1959+650 during the period 2016--2017, until February 2021
using the X-ray data from AstroSAT and Swift and found that the synchrotron peak shifts significantly with different flux states.
\cite{Shah_2021} also used AstroSAT observations to study the anti-correlation between the photon index and the X-ray flux 
using a broken power law.

Around the synchrotron peak, SED is curved and can be well described by log parabolic model \cite[e.g.][]{Landau1986, Massaro(2004), Tramacere_2007, Chen_2014, Gaur_2018, Pandey_2018}). It is 
characterized by the peak energy ($E_{p}$), peak luminosity ($L_{p}$) and the curvature parameter $\beta$. 
The log parabolic spectral distribution arises when the acceleration probability is a decreasing function of electron energy.
By analyzing large X-ray observations, it is suggested by  \cite{Massaro(2004), Tramacere_2007, Tramacere_2011} that the observed anti-correlation expected between $E_{p}$ and $\beta$ could be used as a signature of a stochastic component
in the acceleration process.
For the blazar population, an apparent anti-correlation is expected 
between $E_{p}$ and $L_{p}$ \cite[e.g.][]{Tramacere_2007, Massaro(2008), Kapanadze2016, Kapanadze_2017, Kapanadze2018a} which might 
be associated with the change in average electron energy, beaming factor or magnetic field \cite[e.g.][]{Tramacere_2009}.

In the present work, we analyzed Swift observations of 1ES 1959+650 during the period June 2018$-$December 2020 in 125 nights. We also studied two XMM--Newton satellite observations available during this period. Our aim is to study the temporal/spectral variability of this source on timescales from minutes to years covering different flux states. Studying flux and spectral variability during different flux states are important as distinct physical processes may play a dominant role in different flux states. Flux and spectral variations of blazars on diverse timescales arises either due to pure intrinsic phenomenon such as the 
interaction of relativistic shocks with particle density or magnetic field irregularities in the jet \cite[e.g.][]{Marscher_2014}; production of 
minijets-in-a-jet \cite[e.g.][]{Giannios_2009} or due to extrinsic mechanisms.  Extrinsic mechanisms involve the geometrical effects that 
results due to bending of the jets, either through instabilities \cite[e.g.][]{Pollack_2016} or through orbital motion \cite[e.g.][]{Fromm_2013, Larionov_2020, Valtonen_Wiik_2012}. Long term variability is generally attributed to a mixture of intrinsic as well as extrinsic mechanisms which includes shocks propagating down twisted 
jets or relativistic plasma blobs moving downstream helical structure in the magnetized jets \cite[e.g.][]{Marscher_2008}. Spectral energy distribution of blazars are explained by leptonic and hadronic models \citep{Bottcher2013a} and flux and spectral variability can be explained by merely changing the SED parameters adopting a common set of physical mechanisms \cite[e.g.][]{Patel_2018, Prince_2019, Ghosal_2022}. 
We fit the spectra using log parabolic model and derived spectral parameters i.e. $E_{p}$,  $L_{p}$ and $\beta$ which varies with different flux states of the source.
 In this work, we analyzed correlations between these spectral parameters during different flux states of the source which could provide tight observational constraints 
upon the acceleration and injection processes of the emitting electrons.  

The paper is structured as follows. Section \ref{sec:2} describes observation and data analysis procedures of Swift and XMM--Newton satellites.
Section \ref{sec:3} provides the analysis techniques used to quantify variability, variability timescales and power spectral density.
Section \ref{sec:4} provides the spectral analysis and various models used in the studies.
In section \ref{sec:5}, we describe the results and their interpretation. Results are discussed and summarized in section \ref{sec:6}.  

\section{Data Reduction and Analysis}
\label{sec:2}
\begin{table}
\scriptsize
\setlength{\tabcolsep}{0.035in}
\centering
\caption {Observation log of XMM--Newton and Swift-XRT data for blazar 1ES 1959+650.} 
\label{tab:fvar}
\begin{tabular}{llcc} \hline\hline
{Satellite}&{Date} {(yyyy mm dd)} &{Exposure} & {F$_{{\textit{var}}}$}    \\
&{(Observation ID)}& {Time (sec)}&   {(\%)}   \\
\hline
XMM-Newton&2019-07-05& 42139.94 & 1.95 $\pm$ 0.07\\ 
&(0850980101) & &\\ 
&2020-07-16 & 31239.34 & 3.12 $\pm$ 0.09 \\
&(0870210101) & &\\ 
\hline 
Swift-XRT&2018-06-03& 1007.94 & - \\
&(00094153007) & &\\ 
&2018-06-10& 988.12 & 1.80 $\pm$ 2.03 \\
&(00094153008) & &\\ 
&2018-06-10& 1002.94 & 1.39 $\pm$ 2.74 \\
&(00034588142) & &\\ 
&2018-06-14 & 1003.10 & - \\
&(00034588143) & &\\ 
\hline\hline
\end{tabular}
\#Complete table of all observations appear in online supplementary material

\end{table}

\subsection{SWIFT}
Neil
Gehrels Swift observatory is a multi-wavelength facility equipped with X-ray Telescope (XRT), the Burst Alert Telescope (BAT) and Ultraviolet/ Optical Telescope (UVOT) \citep{Gehrels_2004}. We retrieved the Swift-XRT \citep{Burrows_2005} data from publicly available HEASARC data archive.  The data reduction is performed with the XRT Data Analysis Software (v.3.6.0) which is a part of the HEASoft package (v.6.28). Level 1 unscreened event files were reduced, calibrated and cleaned with the use of $\tt XRTPIPELINE$ script (version 0.13.5). Latest calibration files of Swift CALDB are used with remote access \footnote {\url{https://heasarc.gsfc.nasa.gov/FTP/caldb}}. All the observations are in Windowed Timing mode. Events are selected with standard filtering criteria of 0-2 grades for Windowed Timing (WT) observations. The sources with its centre pixels lying inside the two pixel radius of bad pixels are not used in the analysis. Source region is extracted with a circular region of 20 pixel radius.
For the Windowed timing mode, the source should appear in the middle of the 1-D image, therefore the background can be selected in regions on either side of the source.

$\tt XRTPRODUCTS$ is used to obtain the light curve and spectrum of the source and background. The obtained source light curve is corrected for the resultant loss of effective area, bad/hot pixels and vignetting with the use of $\tt XRTLCCORR$ task.
The ancillary response files (ARFs) were created using the $\tt XRTMKARF$ task. Source spectra were binned to ensure at least 20 count per bin in order to use the $\chi^{2}$ fitting method.

\subsection{XMM-Newton}
Blazar 1ES 1959+650 is observed on 5th July 2019 and 16th July 2020 by XMM-Newton satellite \citep{Jansen_2001}. We used the European Photon Imaging
Camera (EPIC) pn instrument data. EPIC-pn is most sensitive and less affected by the photon pile-up
effects \citep{Struder_2001}. Data reduction is performed with the use of XMM-Newton Science Analysis System (SAS)
~for the LC extraction.
 We extracted the high energy \mbox{(10 keV $<$ E $<$ 12
keV)} light curve for the full frame of the exposed CCD in order to identify flaring particle background.
We restrict our analysis to the 0.3--10 keV energy range, as~data below 0.3 keV are markedly contaminated by noise events and data above 10 keV are usually dominated
by background flares. Source region is extracted using a circle of 40 arcsec radius centered on the source. Background light curve is obtained from the region that corresponds to
circular annulus centered on the source with inner and outer radius of 50 arcsec and 60 arcsec, respectively. Pile up effect is examined using the SAS task $\tt EPATPLOT$.
We found a pile-up in our observations. In order to remove the pile-up, the central 7.5 arcsec radius region is removed while extracting the LC. Source LCs are obtained for the 0.3-10 keV 
energy band (corrected for background flux and given in units of counts~s$^{-1}$), sampled with a fixed bin size of 0.5 ks.

\begin{figure*}
\centering
\includegraphics[width=5.8cm, angle=0]{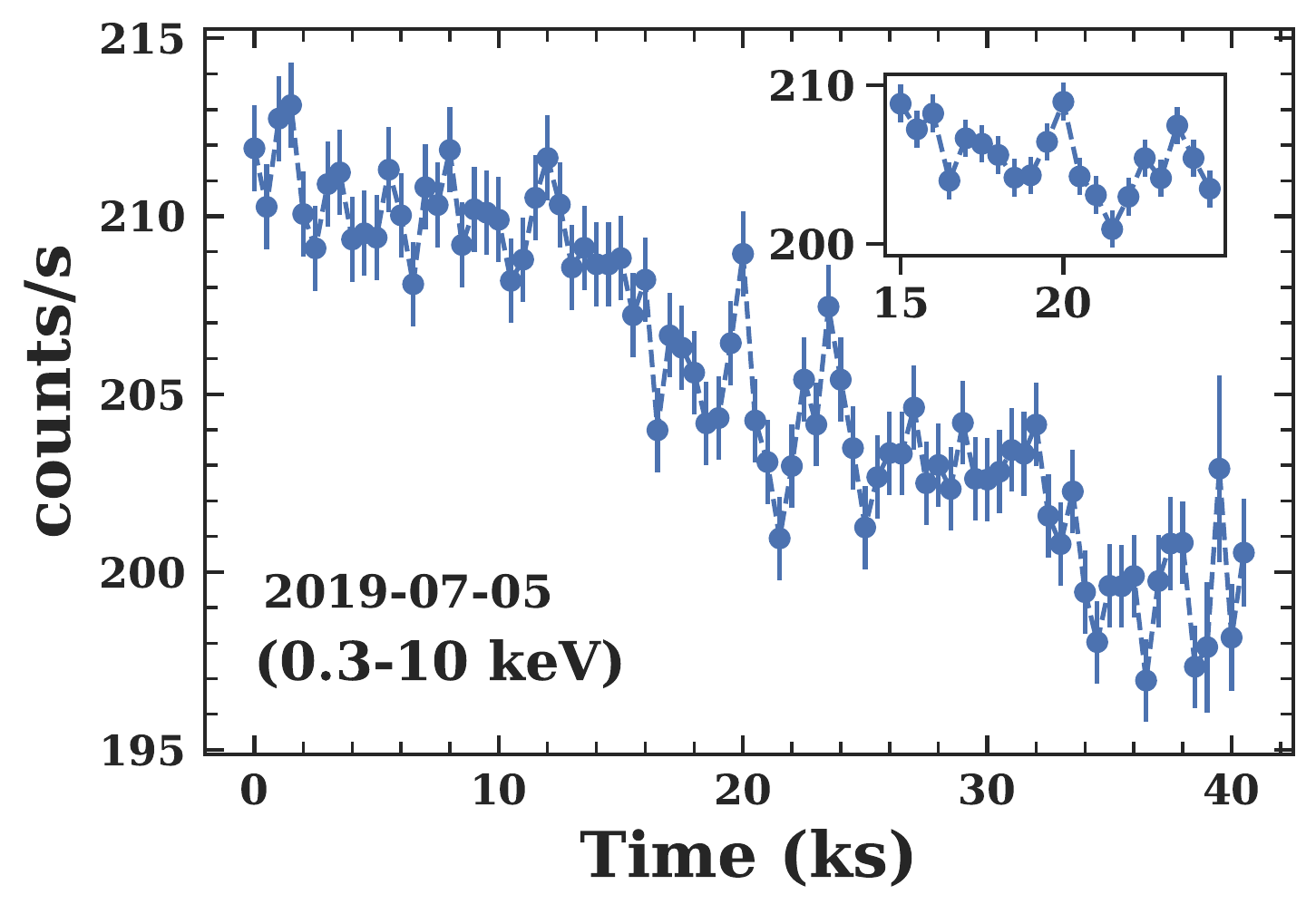}
\includegraphics[width=5.8cm, angle=0]{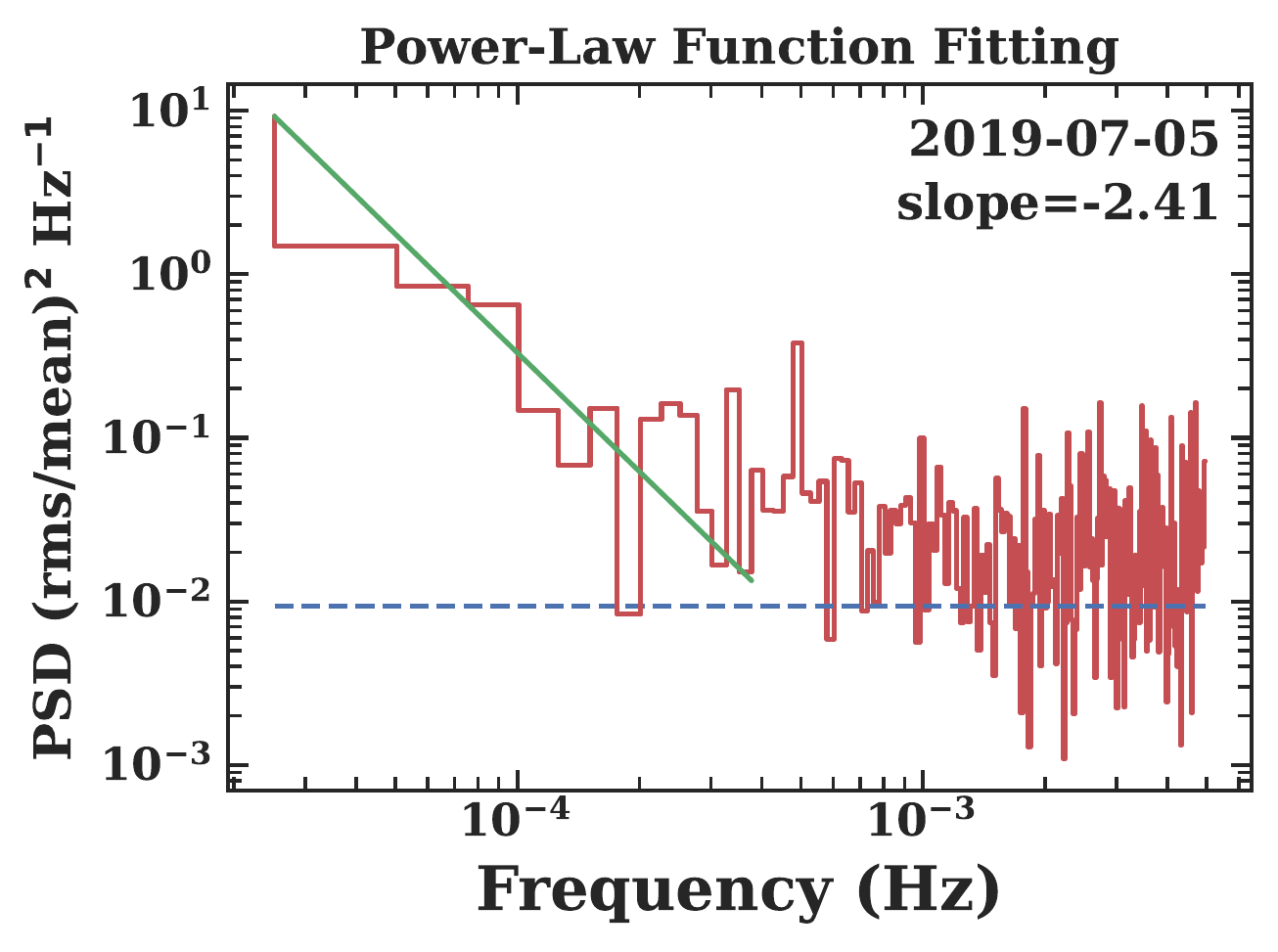}
\includegraphics[width=5.8cm, angle=0]{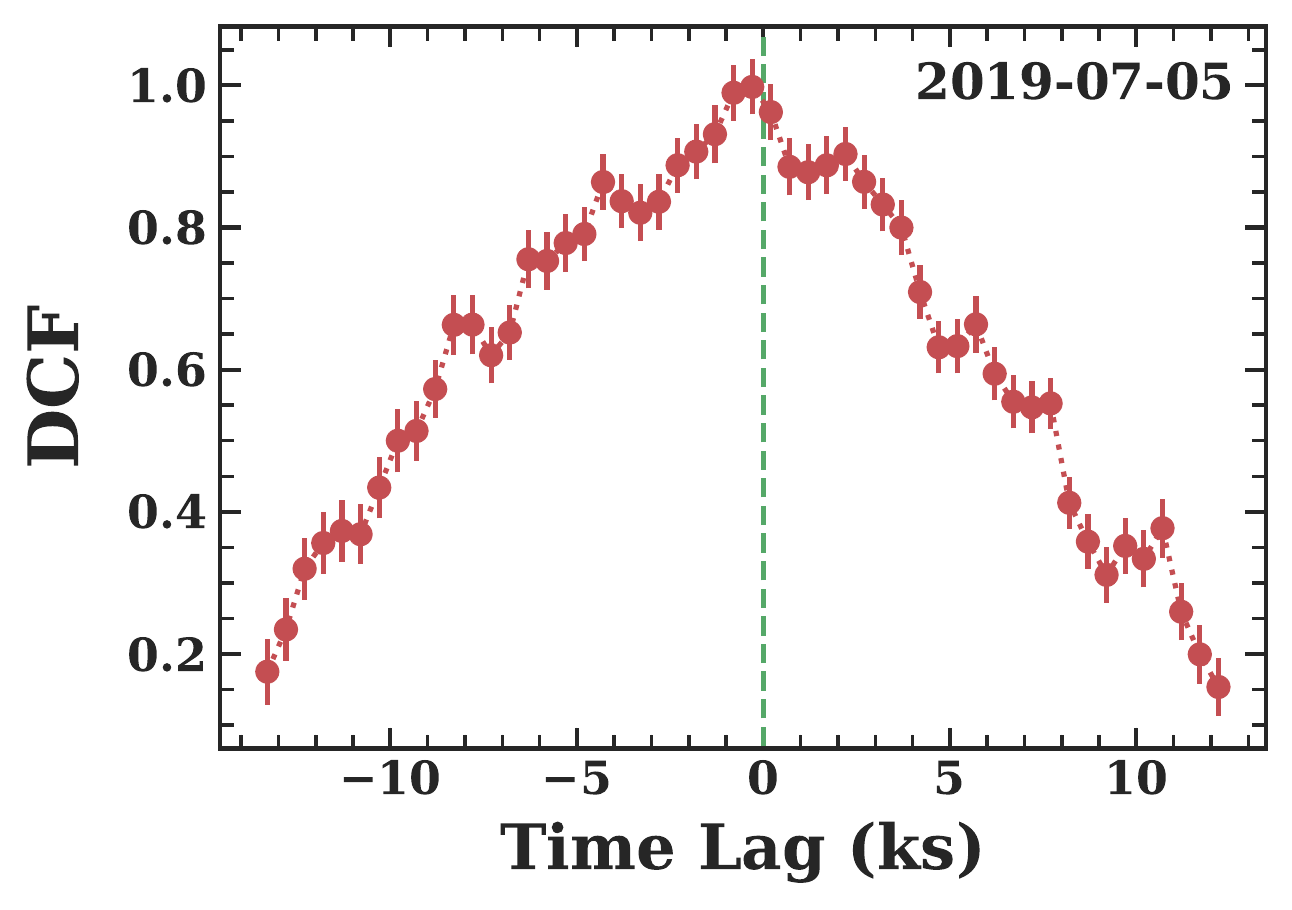}
\includegraphics[width=5.8cm, angle=0]{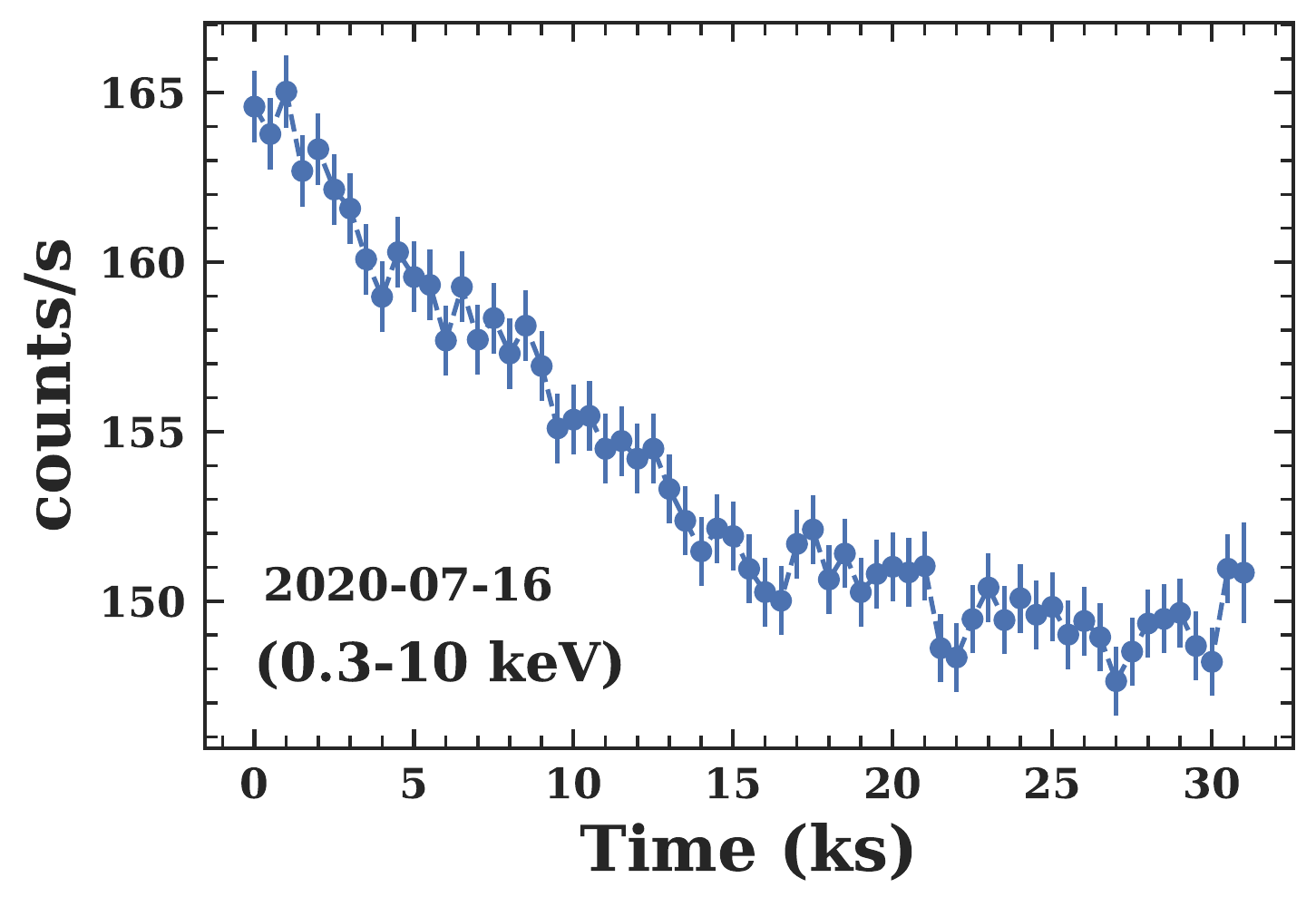}
\includegraphics[width=5.8cm, angle=0]{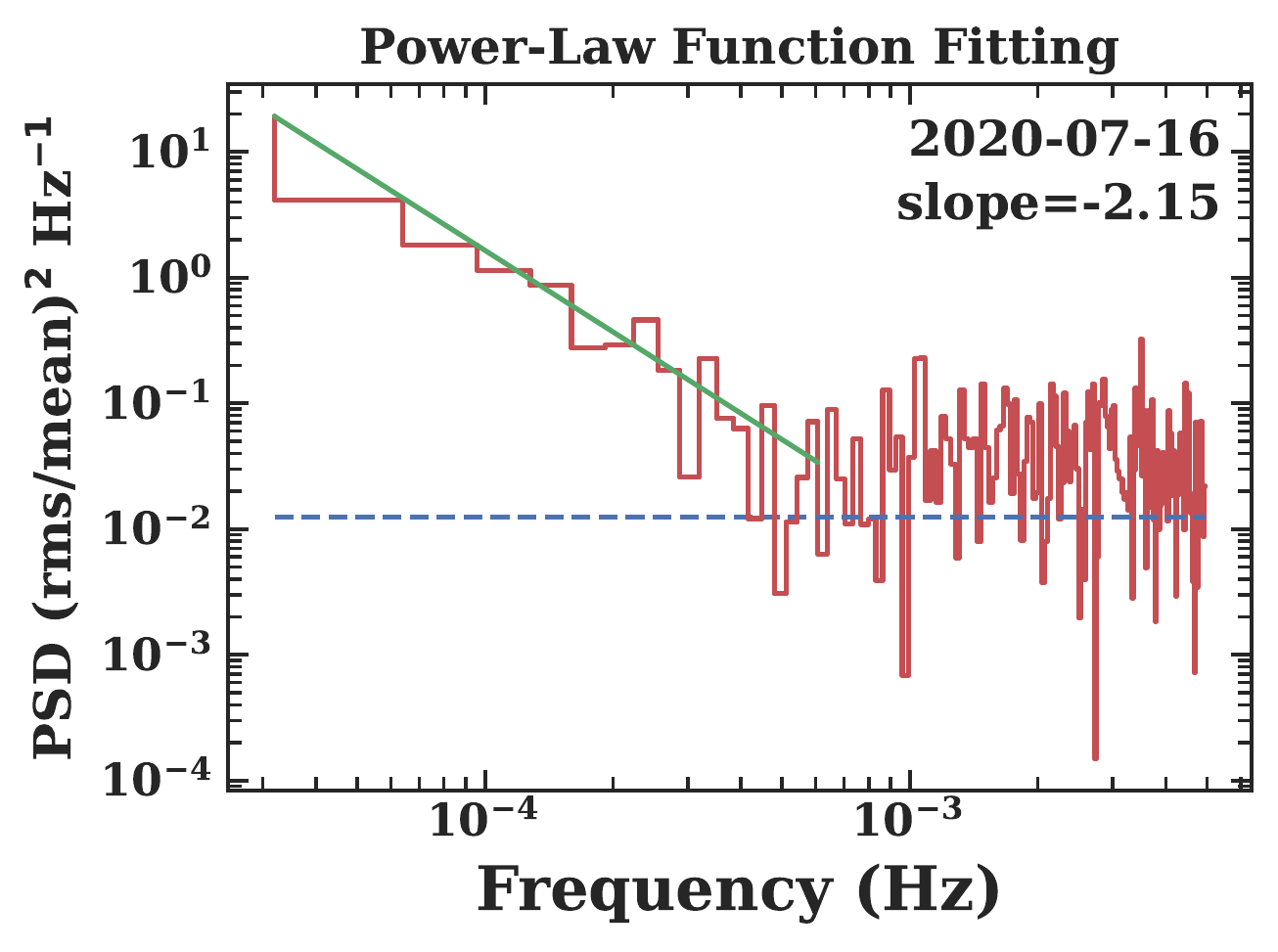}
\includegraphics[width=5.8cm, angle=0]{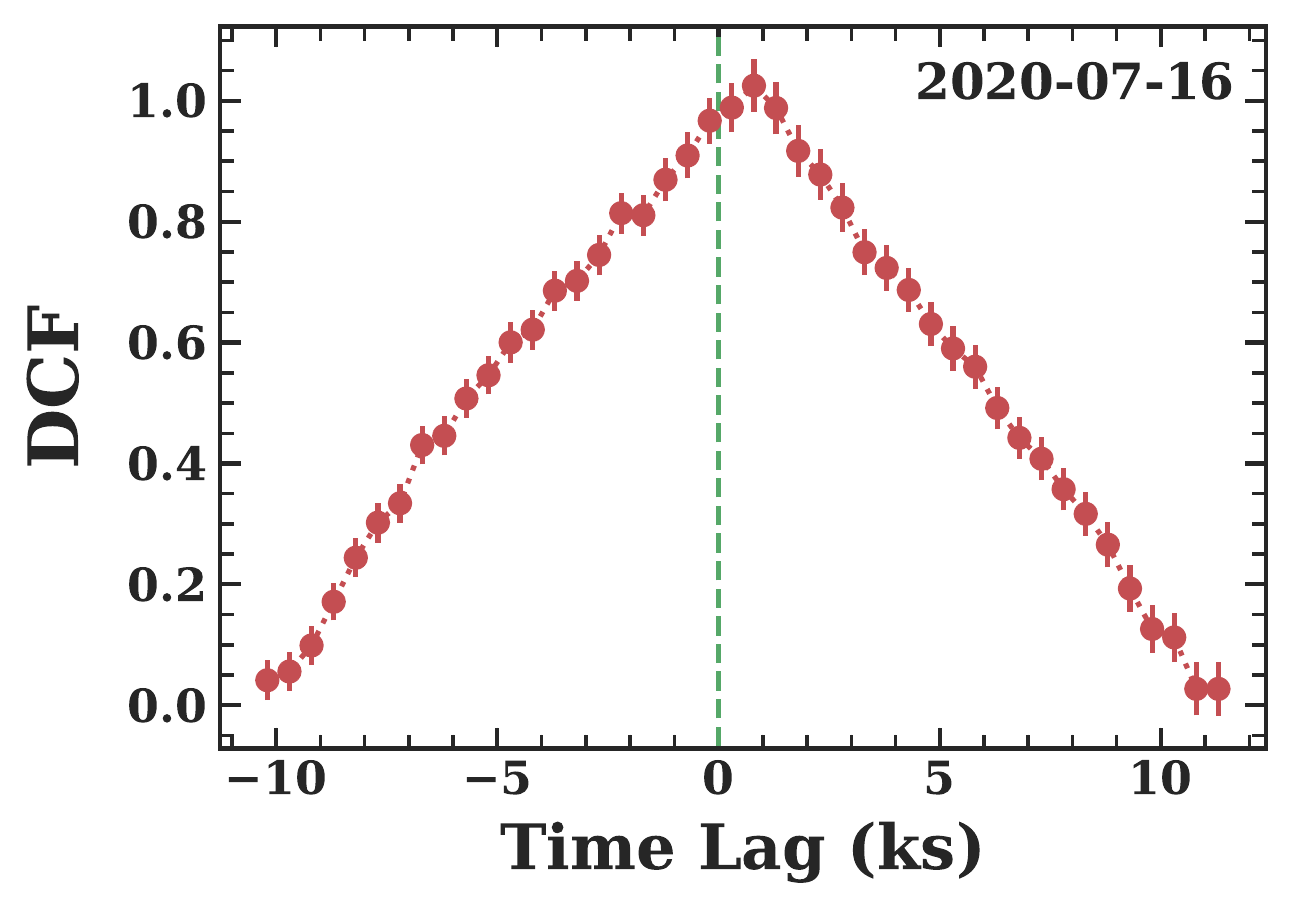}
\caption{XMM-Newton light curves with their corresponding PSD and DCF of Blazar 1ES 1959+650. DCF is performed between X-ray energy 
range of 0.3-2 keV (soft band) and 2-10 keV (hard band). PSD is performed in the X-ray energy range of 0.3--10 keV.}
\label{fig:xmm}
\end{figure*}

\begin{figure*}
\centering
\includegraphics[width=5.8cm , angle=0]{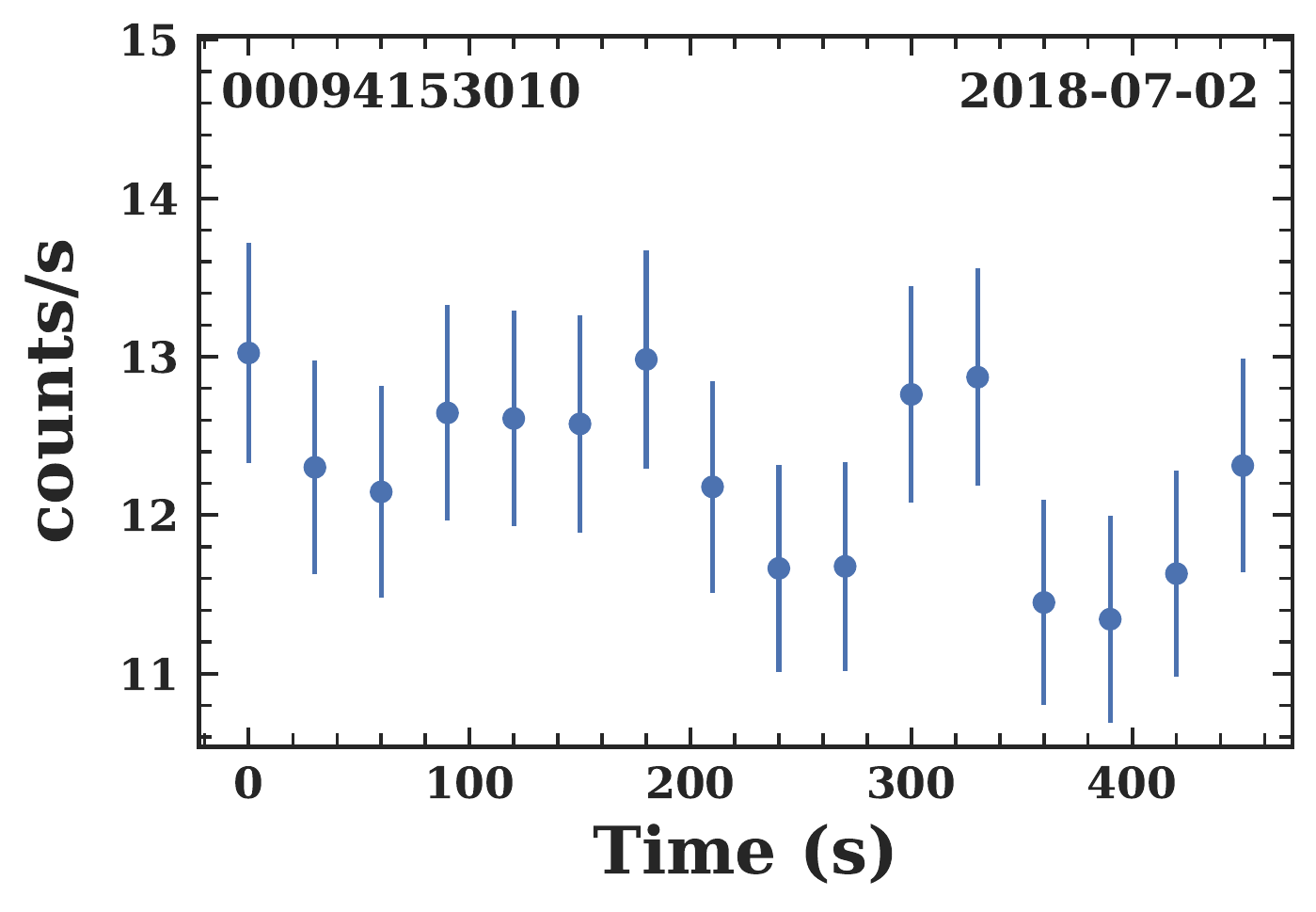}
\includegraphics[width=5.9cm , angle=0]{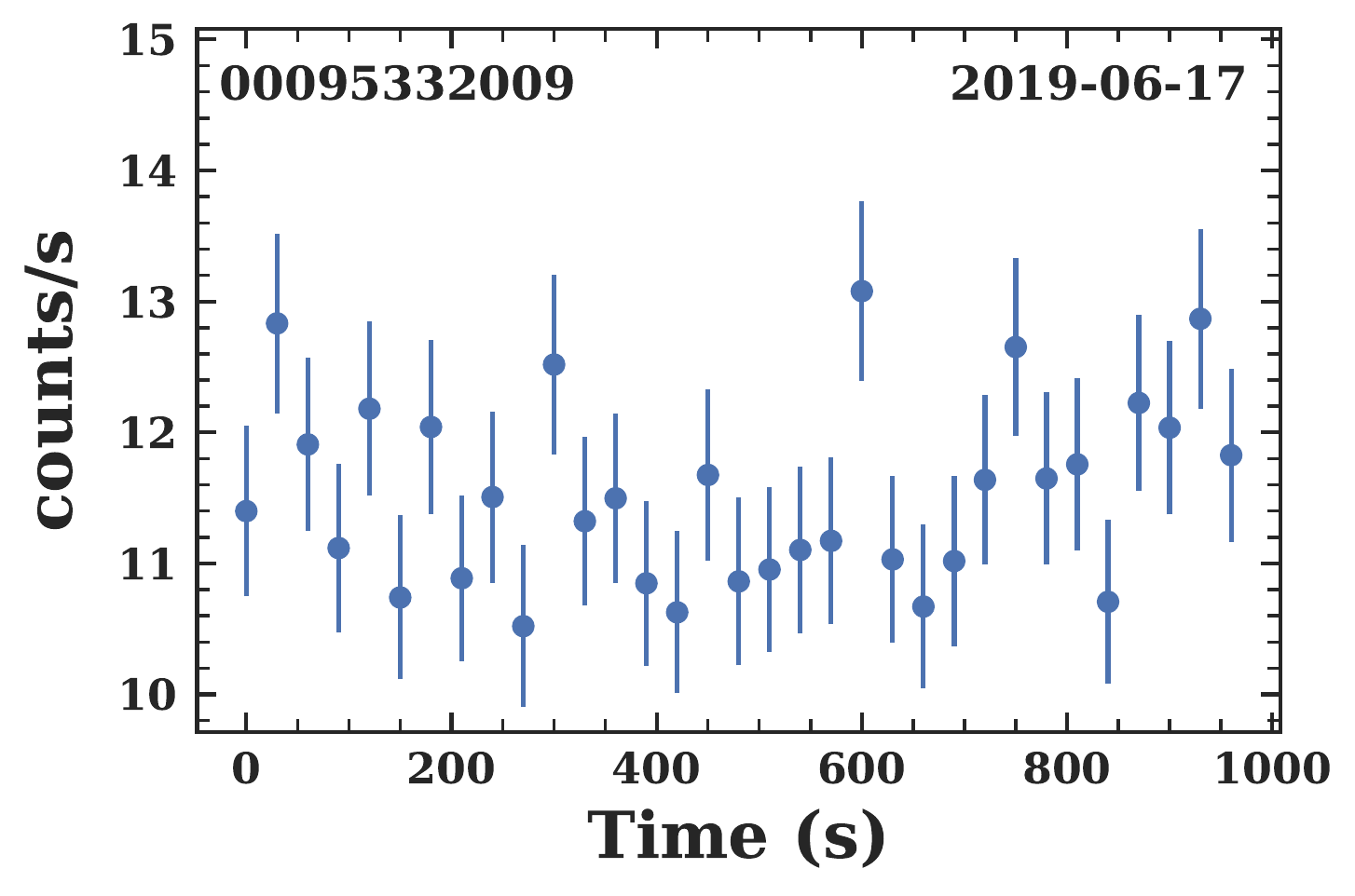}
\includegraphics[width=5.8cm , angle=0]{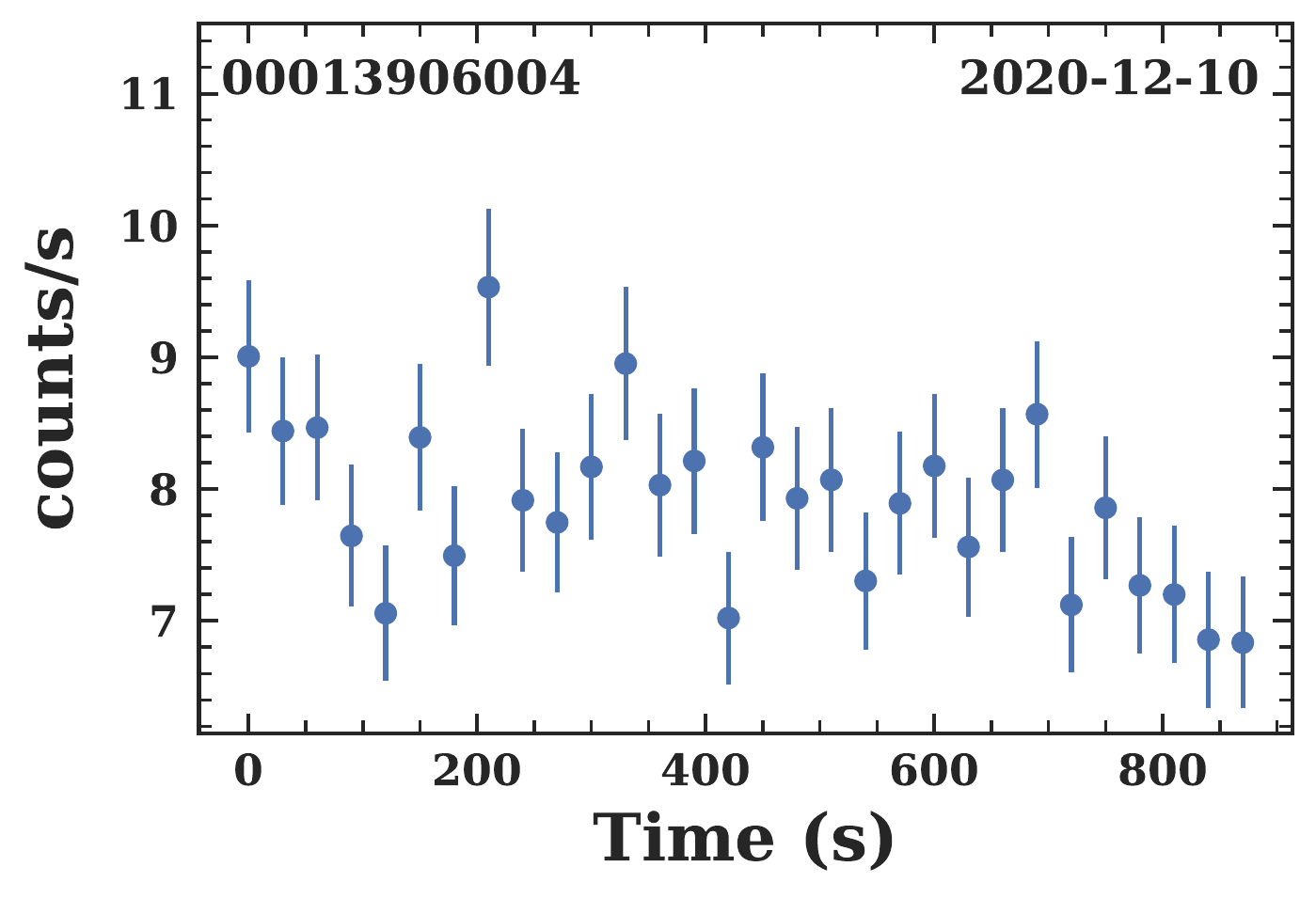}
\includegraphics[width=5.9cm , angle=0]{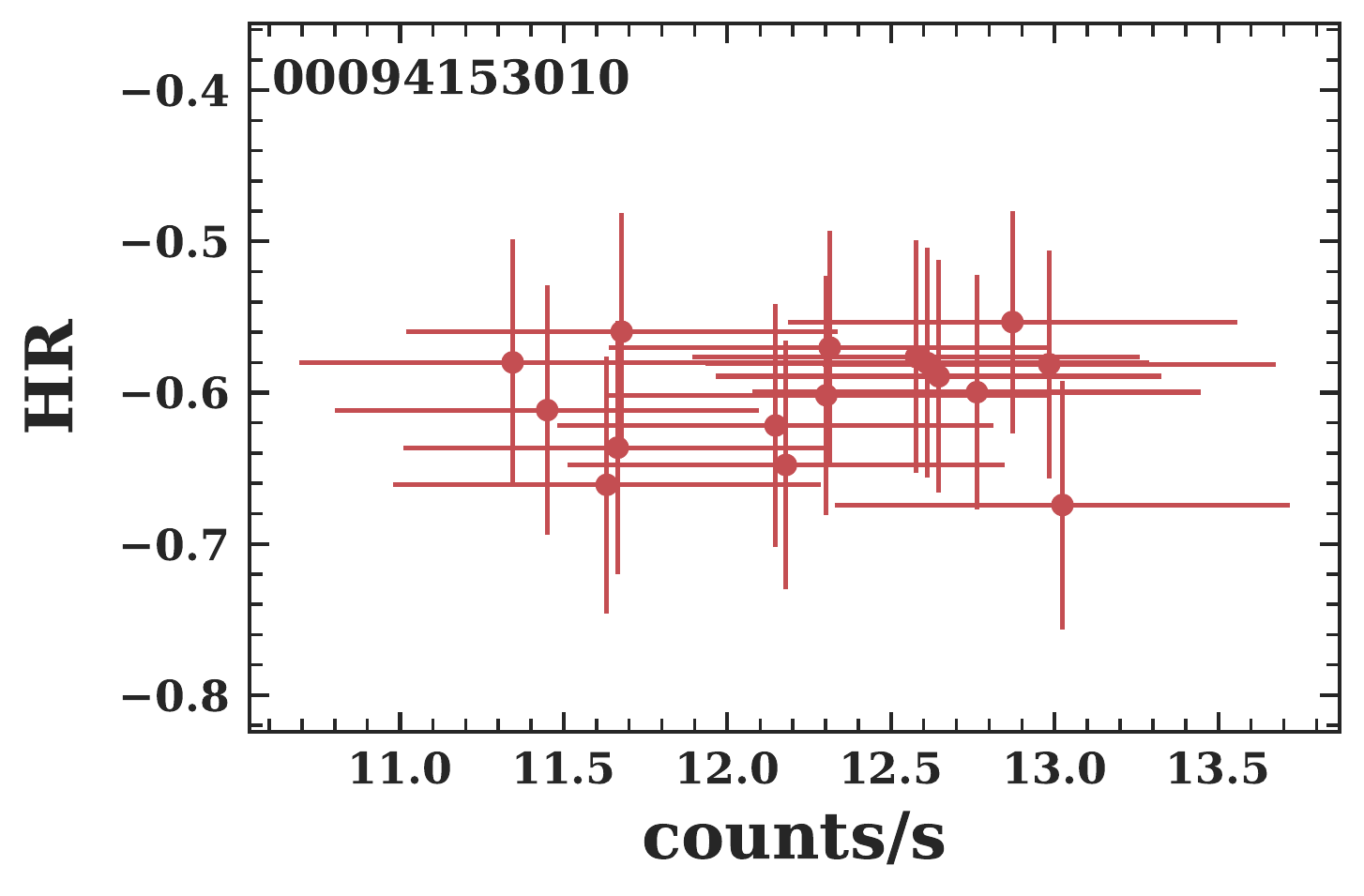}
\includegraphics[width=5.9cm , angle=0]{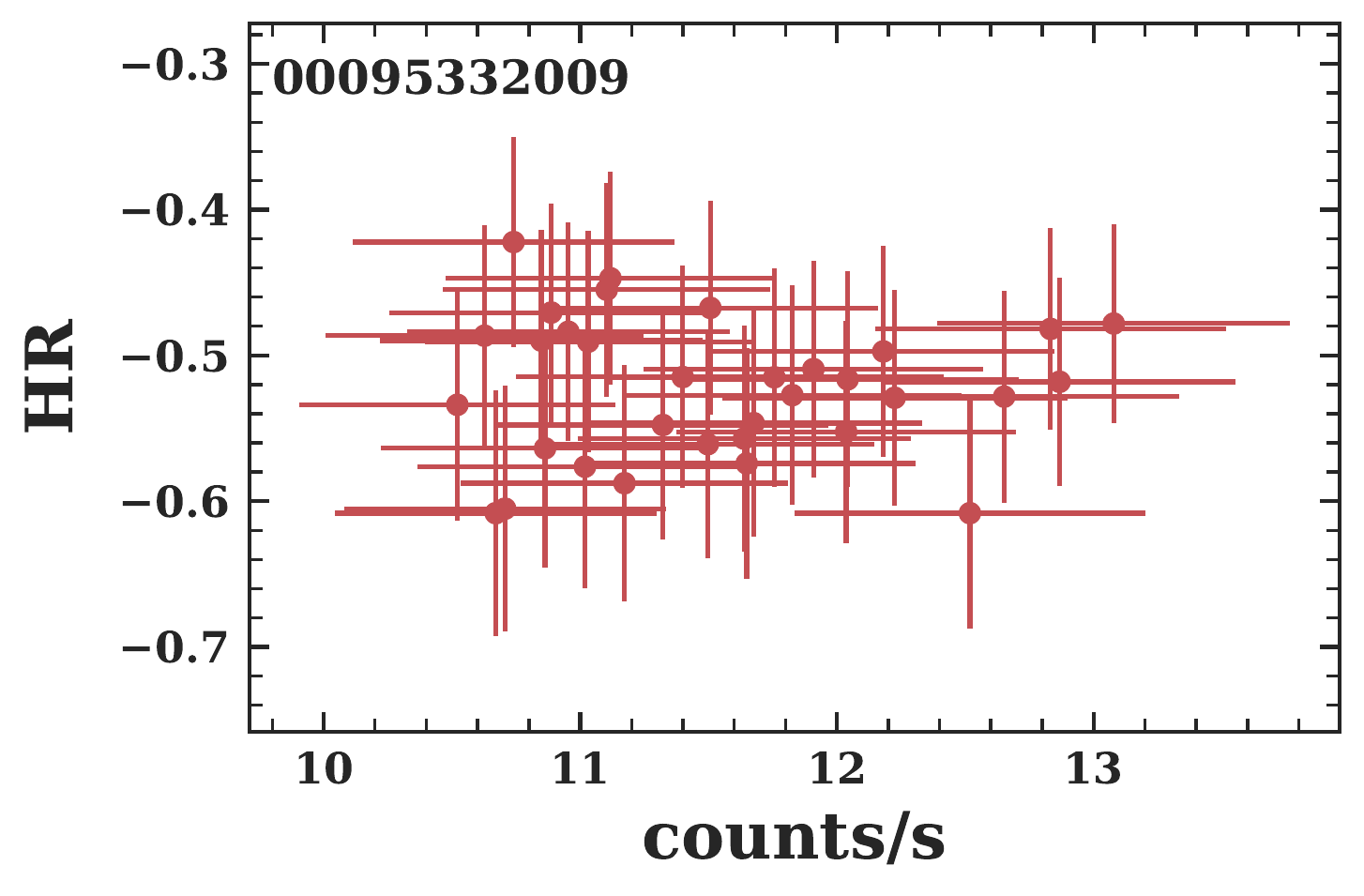}
\includegraphics[width=5.9cm , angle=0]{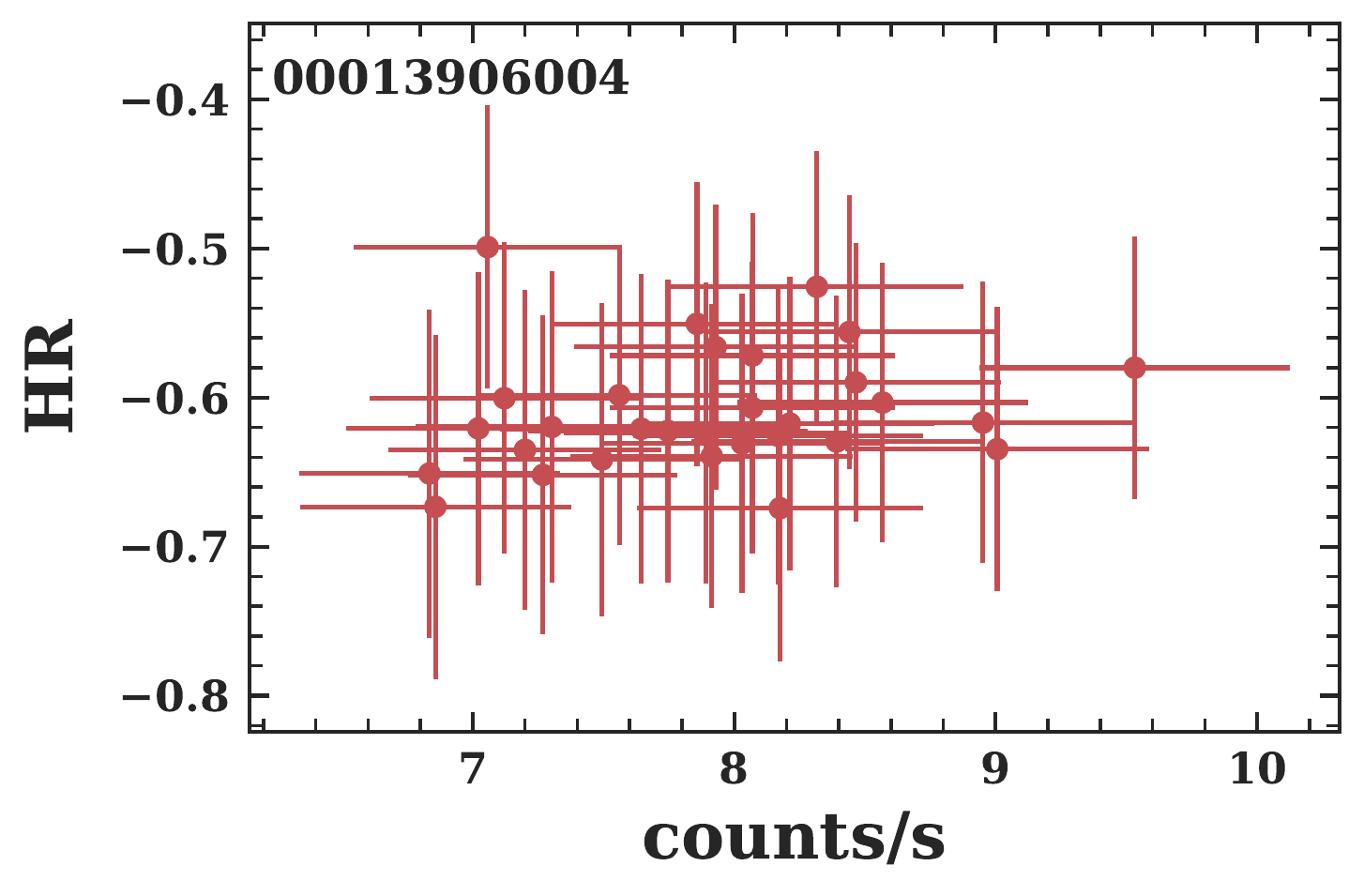}
\caption{Sample of Swift-XRT light curves and their corresponding plot of HR versus counts/s. Hardness Ratio (HR) is calculated between the X-ray energy range of 0.3-2 keV (soft band) and 2-10 keV (hard band). Complete set of light curves of all Swift observations (139 images) appears in a figure set in the online Journal.}
\label{fig:swift}
\end{figure*}

\begin{figure*}
\centering
\includegraphics[width=17cm , angle=0]{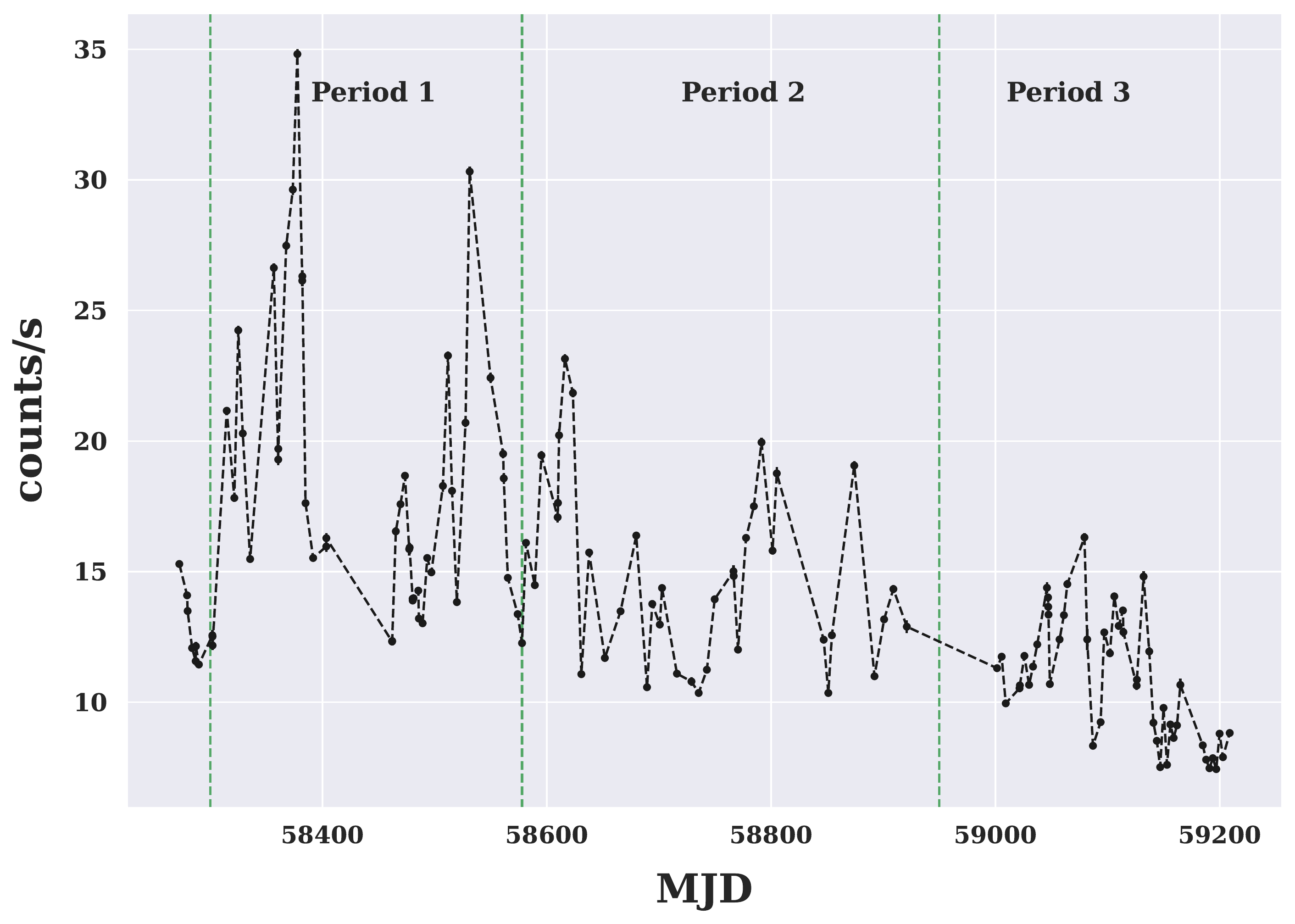}
\caption{Long term light curve of blazar 1ES 1959+650 using Swift-XRT observations from June 2018 to December 2020. Vertical green dashed 
line separates the three periods considered for the analysis. Period 1 corresponds to MJD 58301.75--58577.86; Period 2 corresponds to MJD
58581.37--58920.97 and Period 3 corresponds to MJD 59001.42--59208.88.}
\label{fig:longterm_lc}
\end{figure*}

\section{Analysis Techniques}
\label{sec:3}
\subsection{Excess Variance}
Excess variance ($\sigma^{\:2}_{XS}$) is a measure of source intrinsic variance \citep{Edelson2002, Vaughan(2003a)} evaluated by subtracting the variance that 
arises from measurement errors ($\bar{\sigma}^{\:2}_{err}$) from the total variance of the observed LC ($S^{2}$). If a LC consists of N measured flux values $x_{i}$ with corresponding finite uncertainties $\sigma_{err,\:i}$ arising from measurement errors, then the normalised excess variance ($\sigma^{\:2}_{NXS}$) is calculated as follows:

\begin{equation}
 \sigma^{\:2}_{NXS}= \frac{S^{2} - \bar{\sigma}^{\:2}_{err}}{\bar{x}^{\:2}} \:,
\end{equation}

where $\bar{x}$ is the arithmetic mean of  $x_{i}$, $\bar{\sigma}^{\:2}_{err}= \frac{1}{N} \sum_i\: \sigma^{\:2}_{err,\:i}$ is the mean square error
and $S^{2}$ is the sample variance of the LC, as given by

\begin{equation}
S^{2} = \frac{1}{N-1} \sum_i  (x_{i} - \bar{x})^{2}.
\end{equation}
The fractional rms variability amplitude, $F_{var}$ \citep{EdelsonKrolikPike(1990), Rodriguez-Pascual(1997)}, which is the square root of $\sigma^{\:2}_{NXS}$ is thus

\begin{equation}
F_{var} = \sqrt{\frac{S^{2} - \bar{\sigma}^{\:2}_{err}}{\bar{x}^{\:2}}}.
\end{equation}

The uncertainty on $F_{var}$ is given by \citep{Vaughan(2003a)}.

\begin{equation}
err(F_{var}) = \sqrt{\Biggl(\sqrt{\frac{1}{2N}} \: \frac{\bar{\sigma}^{\:2}_{err}}{\bar{x}^{\:2} F_{var}}\Biggl)^{2}+\Biggl(\sqrt{\frac{\bar{\sigma}^{\:2}_{err}}{N}} \: \frac{1}{\bar{x}}\Biggl)^{2}}.
\end{equation}

\subsection{Hardness Ratio}

Hardness ratio (HR) is useful to characterise the spectral changes over a broad X-ray energy range \cite[e.g.][]{Park_2006, Sivakoff2004}. The energy range of 
(0.3--2) keV and (2--10) keV are used here as soft and hard bands respectively. Hardness ratio is then defined as:
\begin{equation}
HR = \frac{(H - S)}{(H + S)}\:,
\end{equation}
and the error in \emph{HR} ($\sigma_{HR}$) is calculated, as follows:

\begin{equation}
\sigma_{HR} = \frac{2}{(H + S)^{2}} \sqrt{S^{2} \sigma^{\:2}_{H}  + H^{2} \sigma^{\:2}_{S} } \:,
\end{equation}
where \emph{S} and \emph{H} are the net count rates in the soft (0.3--2 keV) and hard (2--10 keV) bands, respectively, while $\sigma_{S}$ and $\sigma_{H}$ are their respective errors \cite[e.g.][]{Pandey2017}.

\subsection{Doubling/Halving Timescales}
Characteristic halving/doubling timescale $\tau$, depending on the
increase or decrease in the ﬂux, is the shortest flux variability time which is calculated from:
\begin{equation}
\label{eq:var_timescale}
    F(t_{1}) = F(t_{2})\: 2^{(t_{1}-t_{2} )/\tau_{var}}\:,
\end{equation}
where $F(t_{1})$ and $F(t_{2})$ are the fluxes of the LC at times $t_{1}$ and $t_{2}$ , respectively. We consider the timescales when the diﬀerence in flux is significant at 3$\sigma$ level \cite[e.g.][]{ Foschini(2011), Dhiman2021}.

\subsection{Discrete Correlation Function}
\label{sec:dcf}
Discrete correlation function is calculated for light curves of two different energy bands. We consider here 0.3-2 keV and 2-10 keV as two energy bands. It is used to investigate a correlation between two unevenly sampled time series data. For such two discrete data sets $a_{i}$ and $b_{j}$, unbinned discrete correlation is defined as \citep{Edelson1988}:

\begin{equation}
    UDCF_{ij}= \frac{(a_{i}-\bar{a}) (b_{i}-\bar{b})}{\sqrt{(\sigma^{2}_{a}-e_{a}^{2}) (\sigma^{2}_{b}-e_{b}^{2}})}\:,
\end{equation}
for all measured pairs ($a_{i}, b_{j}$) with the pairwise lag $\Delta T_{ij} = t_{j}- t_{i}$. $e_{i}$ and $e_{b}, \: \bar{a}$ and $\bar{b} , \: \sigma_{a}$ and $\sigma_{b} $ are the measurement error, mean, standard deviation associated with the data set $a_{i}$ and $b_{j}$ respectively. DCF($\tau$) is obtained by averaging N number of pairs lying in the range $\tau -  \frac{\Delta \tau }{2} < \Delta T_{ij} < \tau + \frac{\Delta \tau }{2}$  where $\Delta \tau$=500s.

\begin{equation}
    DCF(\tau)= \frac{1}{N}\: UDCF_{ij}.
\end{equation}

DCF evaluates the cross-correlation and possible time lags between the soft and hard energy band. The obtained DCF is fitted with the 
Gaussian function \citep{Edelson1988} of the form:

\begin{equation}
\label{eq:dcf_tau}
 DCF(\tau)=a\:\times \:exp \Biggl[\frac{-(\tau - \tau_{lag})^{2}}{(2 \sigma^{2})}\Biggl]\:,
\end{equation}

where ${\tau_{lag}}$ is the time lag at which DCF peaks and $\sigma$ is the width of Gaussian function (used in \citealt{Gaur(2015)}).

\subsection{Power Spectral Density (PSD)}
Power spectral density is the measure of variability power as a function of temporal frequency. It is used to characterize the temporal variations in flux and noise processes in general. Periodogram is a tool to find hidden periodicities including any quasi-periodic oscillations (QPOs). It is defined as the modulus-squared of the Discrete Fourier Transform (DFT) of the data \citep{Vaughan(2005)}. 
The obtained power spectrum consists of red noise which dominates over the measurement error i.e the poisson noise at the lower 
frequencies while the white noise dominates the red noise at higher frequencies which becomes equivalent to the poisson noise level 
of the data. The red noise part of the power spectrum is then fitted with a model of the form $P(f)$= N f$^{\: -m}$ where N is 
the normalisation and $m$ is the power-law spectral index ($m >$ 0) \citep{van1989, Gonz(2012)}. Equations used in determining PSDs are provided in 
Appendix.

\subsection{Spectral Analysis}

The XSPEC software package version 12.11.1 is used for spectral fitting. The Galactic absorption $n_{H}$ is fixed to be 1.0 $\times$ $10^{21}$ cm$^{-2}$ \citep{LockmanSavage(1995)} and the Xspec routine “cflux” is used to obtain unabsorbed flux and its error.

\cite{Massaro(2004), Massaro(2008)} found that 
blazars spectra are curved which arise due to log parabolic electron distributions.
Therefore, they are well described by the log parabola model e.g.,~\citep{Tramacere_2007, Tramacere_2009}.
We~fit each spectra using models which are defined as follows: 
\begin{enumerate}
\item Power law model, which is defined by $k~E^{-\alpha}$.
It is characterized by the photon index $\alpha$, redshift \emph{z}, and Normalization \emph{k}.

\item
The log-parabolic model {\it logpar}. It is characterized by photon index $\alpha$, curvature $\beta$, and Normalization \emph{k}.

\item
Another form of log-parabolic model i.e. {\it eplogpar} model is used to calculate synchrotron peak E$_{p}$. 
Details of the equations are provided in Appendix.

\end{enumerate}

\begin{table}
\scriptsize
\setlength{\tabcolsep}{0.035in}
\centering
\caption{Results of DCF and PSD analysis of 1ES 1959$+$650.}
\label{tab:dcf_psd}
\begin{tabular}{ccccc} \hline\hline
{Observation} & {$\tau_{lag}$(ks)}& $\mathbf{\sigma (ks)}$ & {m} & $\mathbf{log(N)}$\\
{Date}&&&&\\ \hline
2019-07-05&-0.94$\pm$0.10&7.53$\pm$0.12&-2.41$\pm$0.10&-10.13\\
2020-07-16& 0.36$\pm$0.07&5.06$\pm$0.08&-2.15$\pm$0.03&-8.39 \\\hline\hline
\end{tabular}

$\tau_{lag}$ \& $\sigma$ are the time lag at which DCF peaks \& width of the \\Gaussian function fitting at DCF respectively;\\
m \& N are slope \& Normalization constant of the power law \\function fitting at PSD respectively.
\end{table}

\begin{table*}
\centering
\caption{Best spectral fit parameters for the Power Law and Log Parabolic Model of blazar 1ES 1959+650 of XMM-Newton and Swift-XRT Observations from June 2018 to 
December 2020.}
\label{tab:spectral_fitting}
\begin{tabular}{ccccccccccc} \hline\hline
{Observation} &{Fitting}&\boldmath{$\alpha$}& \boldmath{$\beta$} & $\mathbf{log_{10}Flux}$ & \boldmath{$\chi{^{2}_{{Red}}}$}&{DoF}&{F-test}&{p-value}&{E$_{p}$}&{L$_{p}$(10$^{45}$)}\\
{ID}& {Model}&& & {(ergs/cm$^{2}$/s)}&& &&&{(keV)}&{(erg/s)}\\
\hline
&&&&&XMM-Newton&&&&&\\\hline
0850980101&PL&$2.13^{+0.002}_{-0.002}$&-&${-9.16}^{+0.001}_{-0.001}$&41.47&175&-&-&-&-\\
&LP&$2.06^{+0.002}_{-0.002}$&$0.22^{+0.005}_{-0.005}$&${-9.18}^{+0.001}_{-0.001}$&5.57&174&1129.34&5.37e-78&0.74(0.01)&1.20(0.002)\\\hline
0870210101&PL&$1.99^{+0.002}_{-0.002}$&-&${-9.27}^{+0.001}_{-0.001}$&29.67&175&-&-&-&-\\
&LP&$1.89^{+0.003}_{-0.003}$&$0.24^{+0.006}_{-0.006}$&${-9.29}^{+0.001}_{-0.001}$&2.79&174&1687.02&1.83e-91&1.66(0.02)&0.89(0.002)\\\hline

&&&&&SWIFT-XRT&&&&&\\\hline
00094153007&PL&$2.2^{+0.021}_{-0.021}$&-&${-9.21}^{+0.006}_{-0.006}$&1.36&288&-&-&-&-\\
&LP&$2.09^{+0.032}_{-0.033}$&$0.31^{+0.068}_{-0.067}$&${-9.24}^{+0.009}_{-0.009}$&1.14&287&56.59&6.93e-13&0.71(0.13)&1.12(0.02)\\\hline
00094153008&PL&$2.18^{+0.022}_{-0.022}$&-&${-9.25}^{+0.006}_{-0.007}$&1.15&280&-&-&-&-\\
&LP&$2.06^{+0.034}_{-0.035}$&$0.36^{+0.073}_{-0.071}$&${-9.29}^{+0.009}_{-0.009}$&0.88&279&86.25&4.66e-18&0.83(0.12)&1.00(0.02)\\\hline
00034588142&PL&$2.21^{+0.022}_{-0.022}$&-&${-9.28}^{+0.007}_{-0.007}$&1.29&265&-&-&-&-\\
&LP&$2.07^{+0.035}_{-0.036}$&$0.44^{+0.078}_{-0.075}$&${-9.32}^{+0.010}_{-0.010}$&0.90&264&116.88&8.55e-23&0.84(0.10)&0.96(0.02)\\\hline
00034588143&PL&$2.25^{+0.026}_{-0.025}$&-&${-9.31}^{+0.007}_{-0.007}$&1.07&241&-&-&-&-\\
&LP&$2.15^{+0.037}_{-0.038}$&$0.32^{+0.085}_{-0.082}$&${-9.34}^{+0.011}_{-0.011}$&0.89&240&50.52&1.34e-11&0.58(0.15)&0.94(0.03)\\\hline
\hline
\end{tabular}

PL: Power law; LP: Log Parabolic; DoF: Degree of Freedom; E$_{p}$: Peak Energy; L$_{p}$: Peak Luminosity.

Complete table of all observations appear in online supplementary material.
\end{table*}

\begin{table*}
\scriptsize
\setlength{\tabcolsep}{0.035in}
\centering

\caption {Summary of Swift-XRT observations of 1ES 1959$+$650 during different periods.}
\label{tab:counts}
\begin{tabular}{lccccccccc} \hline\hline
& &\multicolumn{2}{c}{{[MJD 58301.75--58577.86]}}& \multicolumn{2}{c}{{[MJD 58581.37–58920.97]}}& \multicolumn{2}{c}{{[MJD 59001.42–59208.88]}}\\
& \multicolumn{2}{c}{{Total Epoch}}& \multicolumn{2}{c}{{Period 1}}& \multicolumn{2}{c}{{Period 2}}& \multicolumn{2}{c}{{Period 3}}\\

\hline\hline\\
Mean counts/s&14.51&&18.28&&14.82&&10.90&\\
Maximum Flux (counts/s)&34.82&&34.82&&23.15&&16.31&\\
$F_{var}$ (\%)&33.97 (0.09)&&29.96(0.14)&&23.12(0.18)&&21.43(0.20)&\\
HR (max$-$min)& -0.42$-$-0.64&&-0.42$-$-0.63 &&-0.46$-$-0.64 && -0.43$-$-0.61&\\
$\alpha$ (max$-$min) & 2.25$-$1.34 &&2.17$-$1.58&&2.25$-$1.56&&2.14$-$1.34&\\
$\beta$ (max$-$min) & 0.99$-$0.21 &&0.84$-$0.31&&0.82$-$0.23&&0.99$-$0.24&\\
$E_{p}$ (keV) (max$-$min)&2.88$-$0.46 &&2.45$-$0.62&&2.28$-$0.46&&2.88$-$0.67&\\
Slope / intercept ($E_{p}$ vs.\ $L_{p}$) &0.47(0.08) /-0.08(0.04)&&0.85(0.08) / 0.06(0.04)&&0.38(0.13) / -0.06(0.06)&&0.45(0.08) / -0.41(0.04)&\\ 
Slope / intercept ($E_{p}$ vs.\ 1/$\beta$) &-0.56(0.16) / 2.35(0.07)&&-0.34(0.2) / 2.03(0.08)&&-0.90(0.41) / 2.50(0.16)&&-0.41(0.30) / 2.50(0.14)&\\
&&&&&&&&\\
& \multicolumn{8}{c}{{Spearman's Correlation Coefficient}}\\
& \multicolumn{2}{c}{\boldmath{$\rho_{s}$} \boldmath{($p$}{-value)}}& \multicolumn{2}{c}{\boldmath{$\rho_{s}$} \boldmath{($p$}{-value)}}& \multicolumn{2}{c}{\boldmath{$\rho_{s}$} \boldmath{($p$}{-value)}}& \multicolumn{2}{c}{\boldmath{$\rho_{s}$} \boldmath{($p$}{-value)}}\\
\cline{2-2}\cline{4-4}\cline{6-6}\cline{8-8}
soft and hard counts/s & 0.90 (7.13e-53) && 0.97 (7.95e-28) && 0.94 (1.79e-18) && 0.95 (3.50e-24)& \\
$Flux_{(0.3-10) keV}$ and HR & 0.41 (5.53e-07) && 0.86 (1.04e-14) && 0.56 (2.98e-04) && 0.64 (8.59e-07)& \\
$Flux_{(0.3-10) keV}$ and $\alpha$ & -0.38(3.08e-06) && -0.73 (6.60e-09) && -0.40 (1.35e-02) && -0.52 (1.51e-04)& \\
$Flux_{(0.3-10) keV}$ and $E_{p}$ & 0.39 (1.51e-06) && 0.79 (5.30e-11) && 0.44 (6.14e-03) && 0.60 (6.10e-06)& \\
$\alpha$ and $\beta$ & -0.48 (2.00e-09) && -0.41 (4.18e-03) && -0.59 (1.05e-04) && -0.48 (6.38e-04)&\\
$E_{p}$ and $\alpha$ & -0.97 (1.73e-87) && -0.99 (6.74e-37) && -0.95 (3.22e-19) && -0.96 (2.15e-27)&\\
$E_{p}$ and 1/$\beta$ & -0.31 (2.04e-04)&& -0.30 (3.84e-02)&& -0.37 (2.36e-02) && -0.26 (7.09e-02)& \\
$E_{p}$ and $L_{p}$ & 0.35 (2.63e-05) && 0.79 (3.08e-11) && 0.41 (1.04e-02) && 0.62 (2.45e-06)&\\
\hline\hline

\end{tabular}\\
max: maximum value; min: minimum value
\label{tab:spearman_coeff}
\end{table*}

\begin{figure*}
\centering
\includegraphics[width=7.0cm , angle=0]{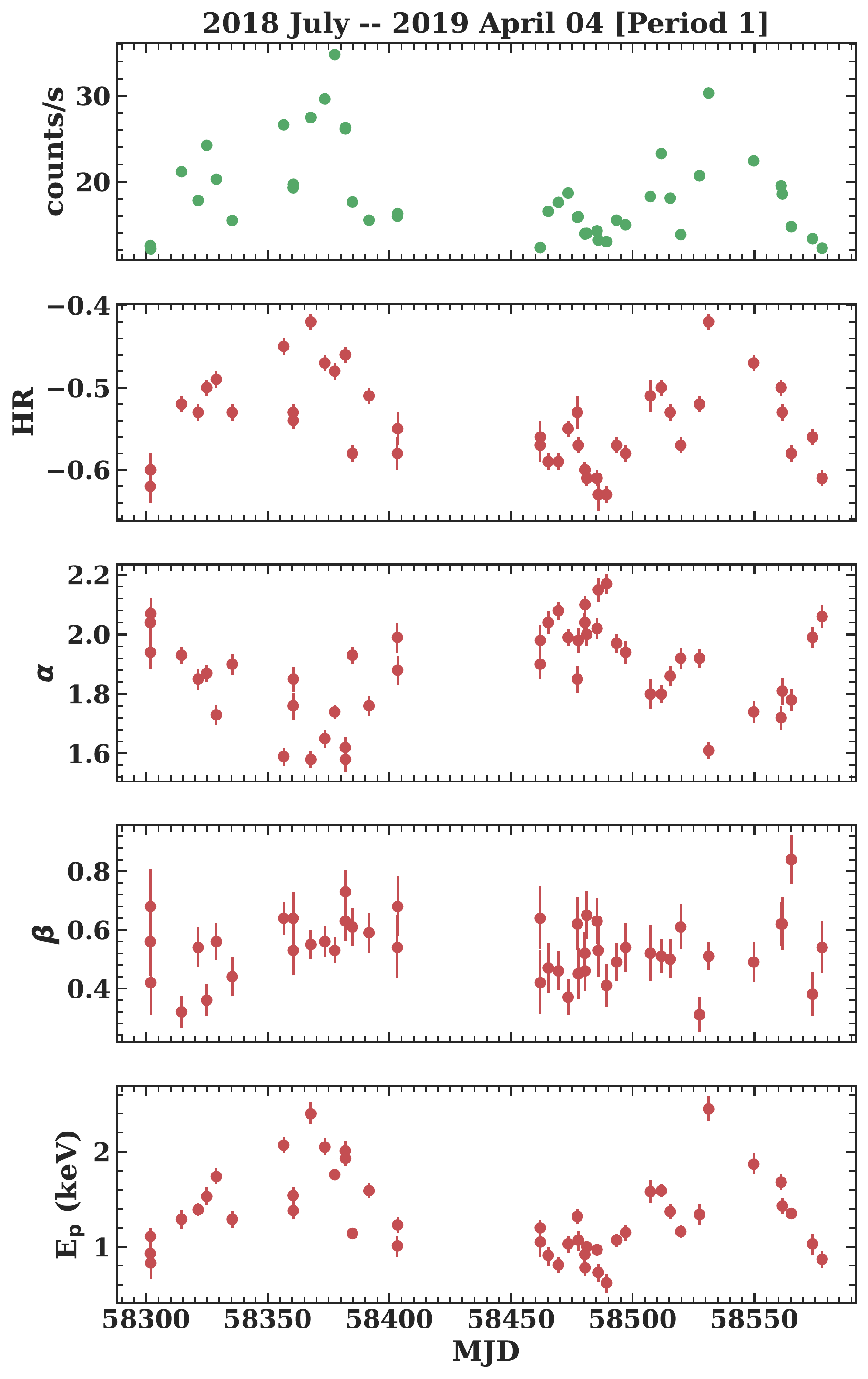}
\includegraphics[width=6.8cm , angle=0]{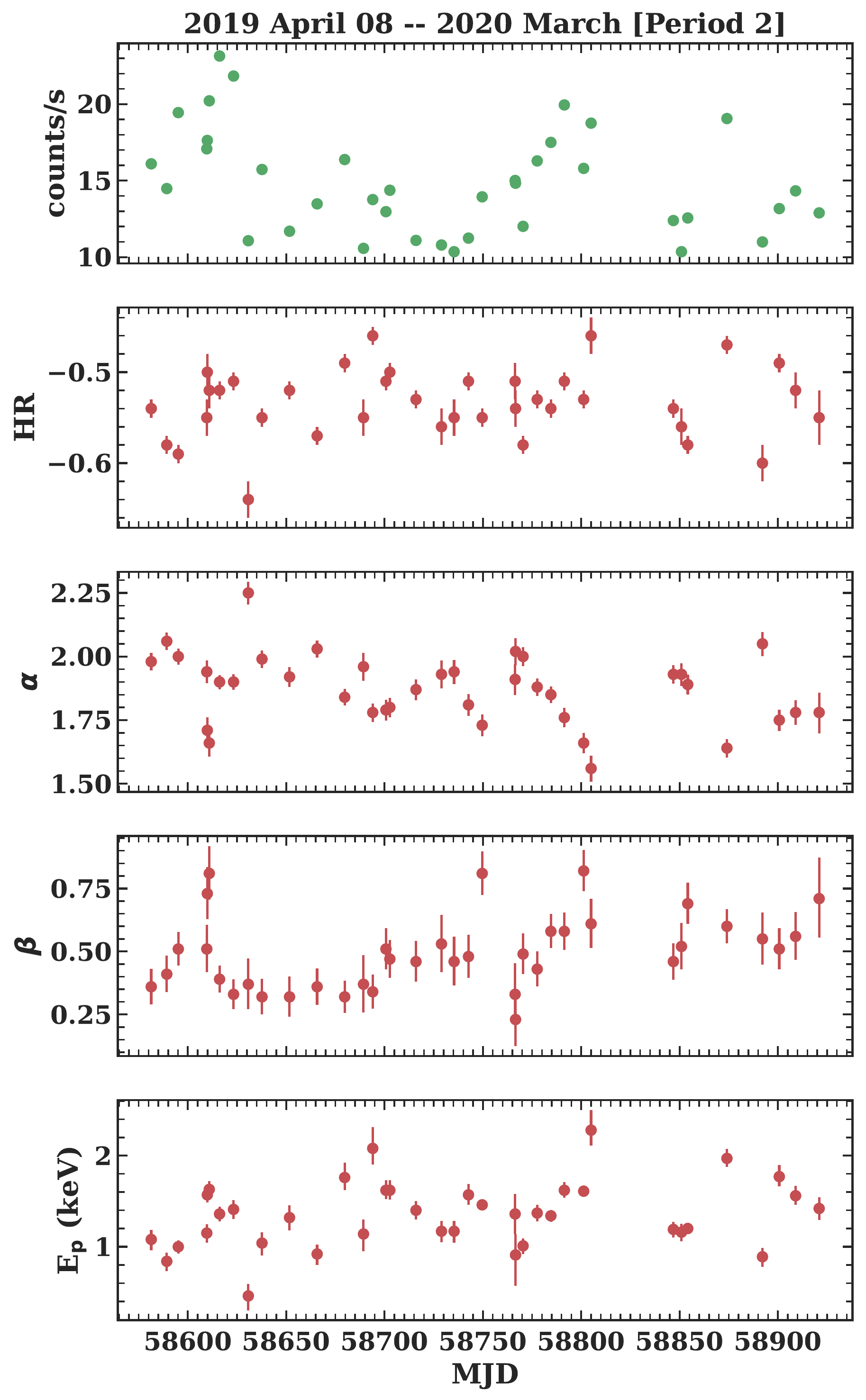}
\includegraphics[width=7.0cm , angle=0]{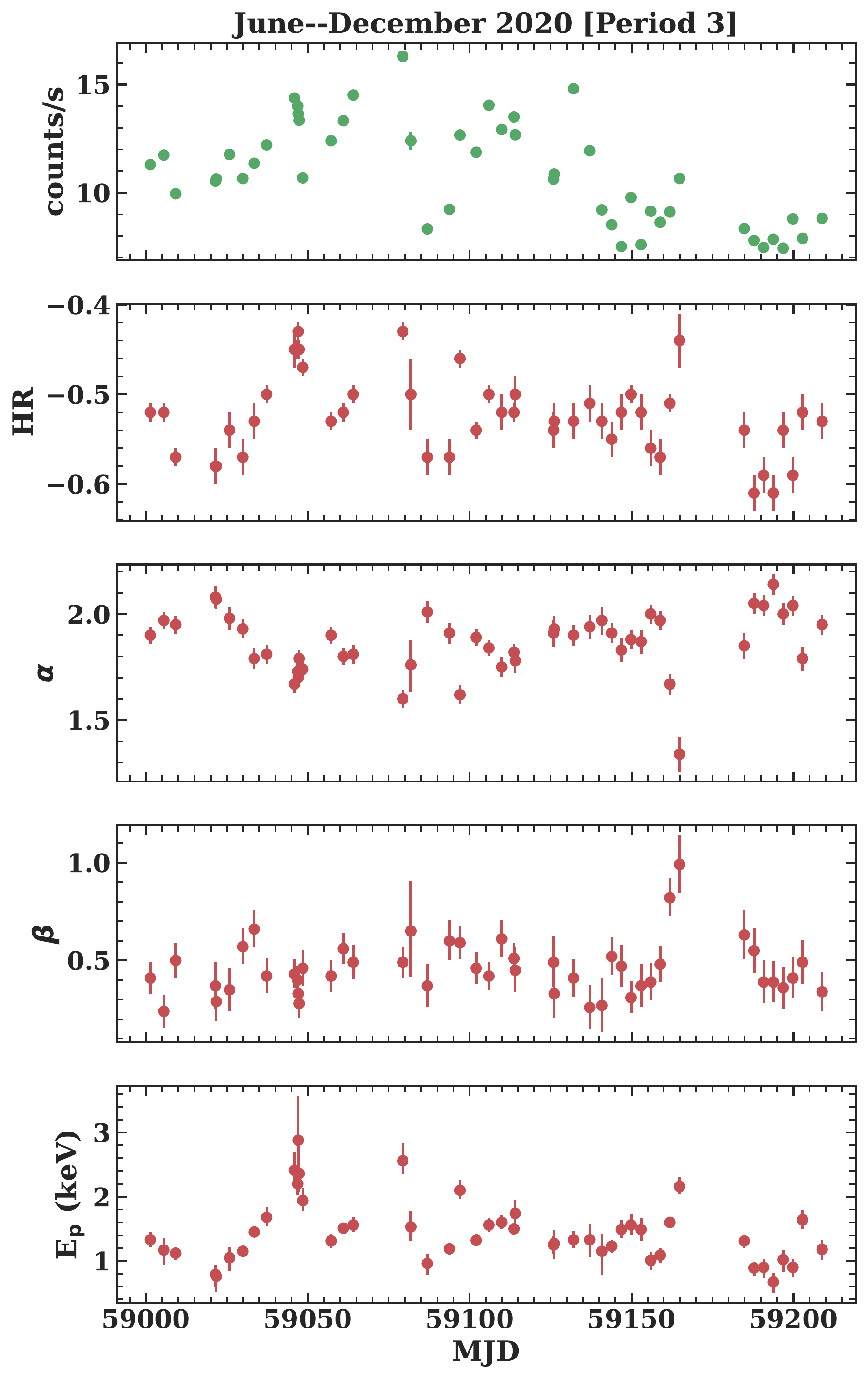}
\caption{Temporal variations of various spectral parameters of blazar 1ES 1959$+$650 during different intervals. Best fit parameters are obtained with 2.7$\sigma$ confidence interval.}
\label{fig:temporal_var}
\end{figure*}

\begin{figure*}
\centering
\includegraphics[width=5.3cm , angle=0]{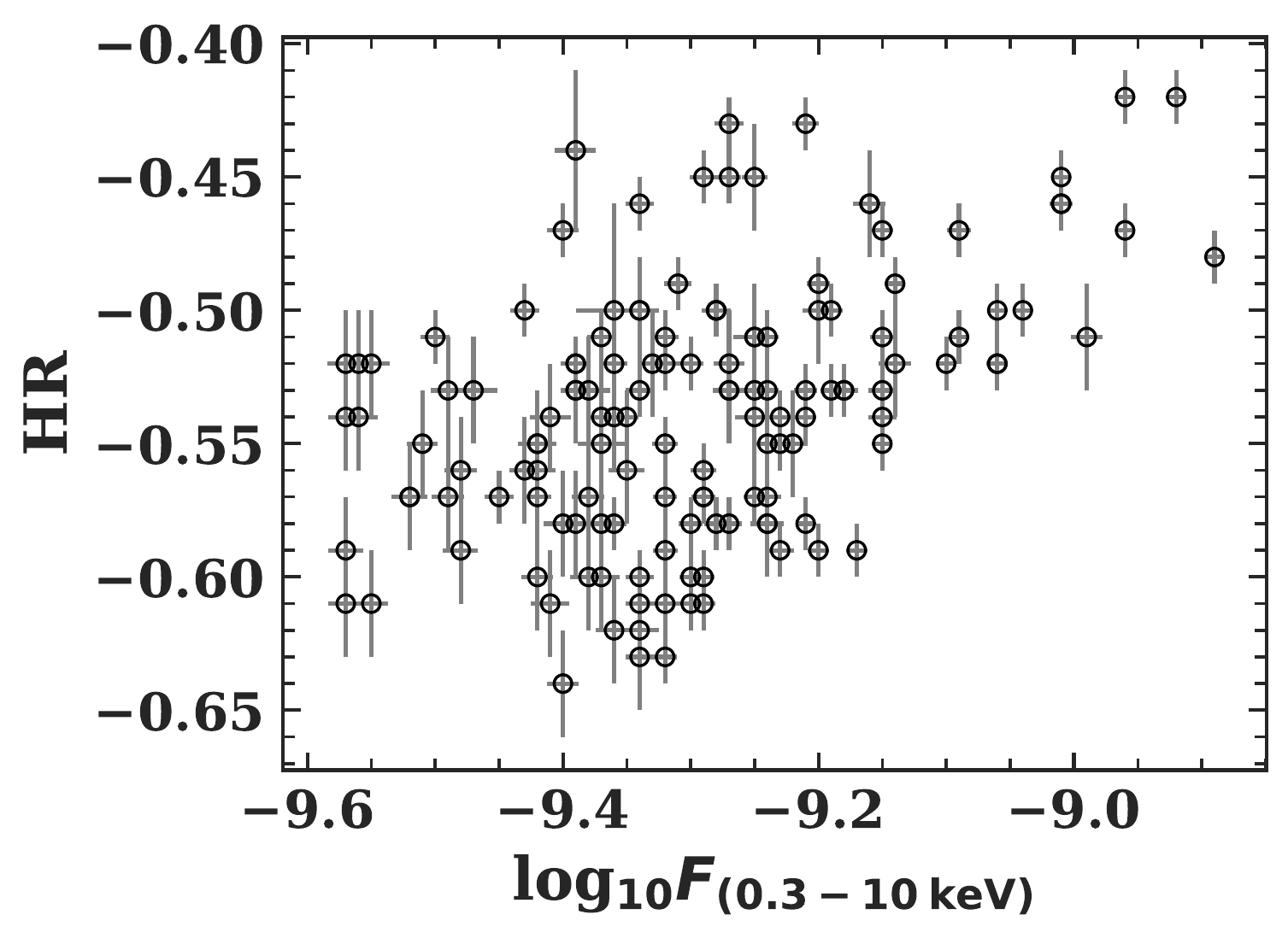}
\includegraphics[width=5cm , angle=0]{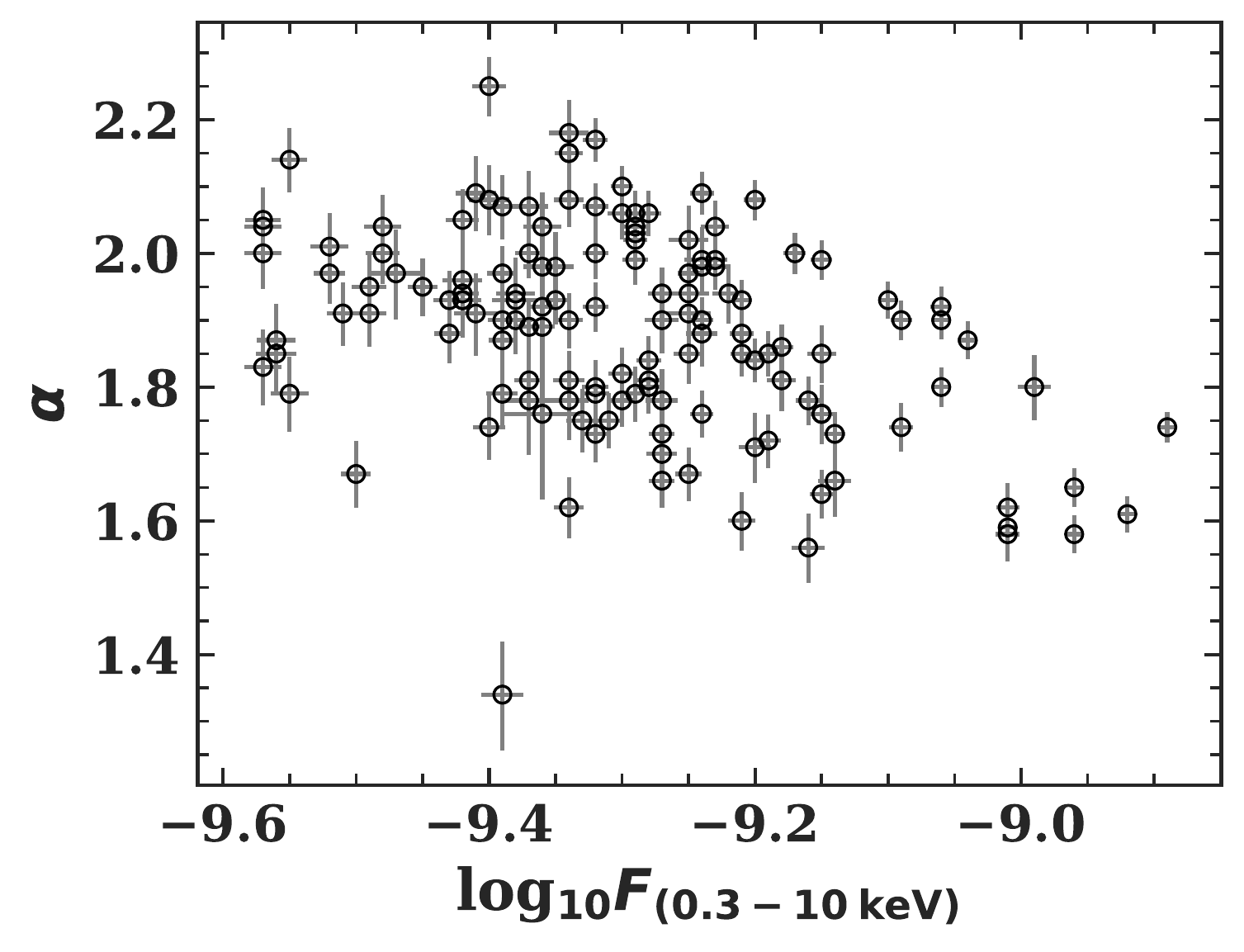}
\includegraphics[width=5cm , angle=0]{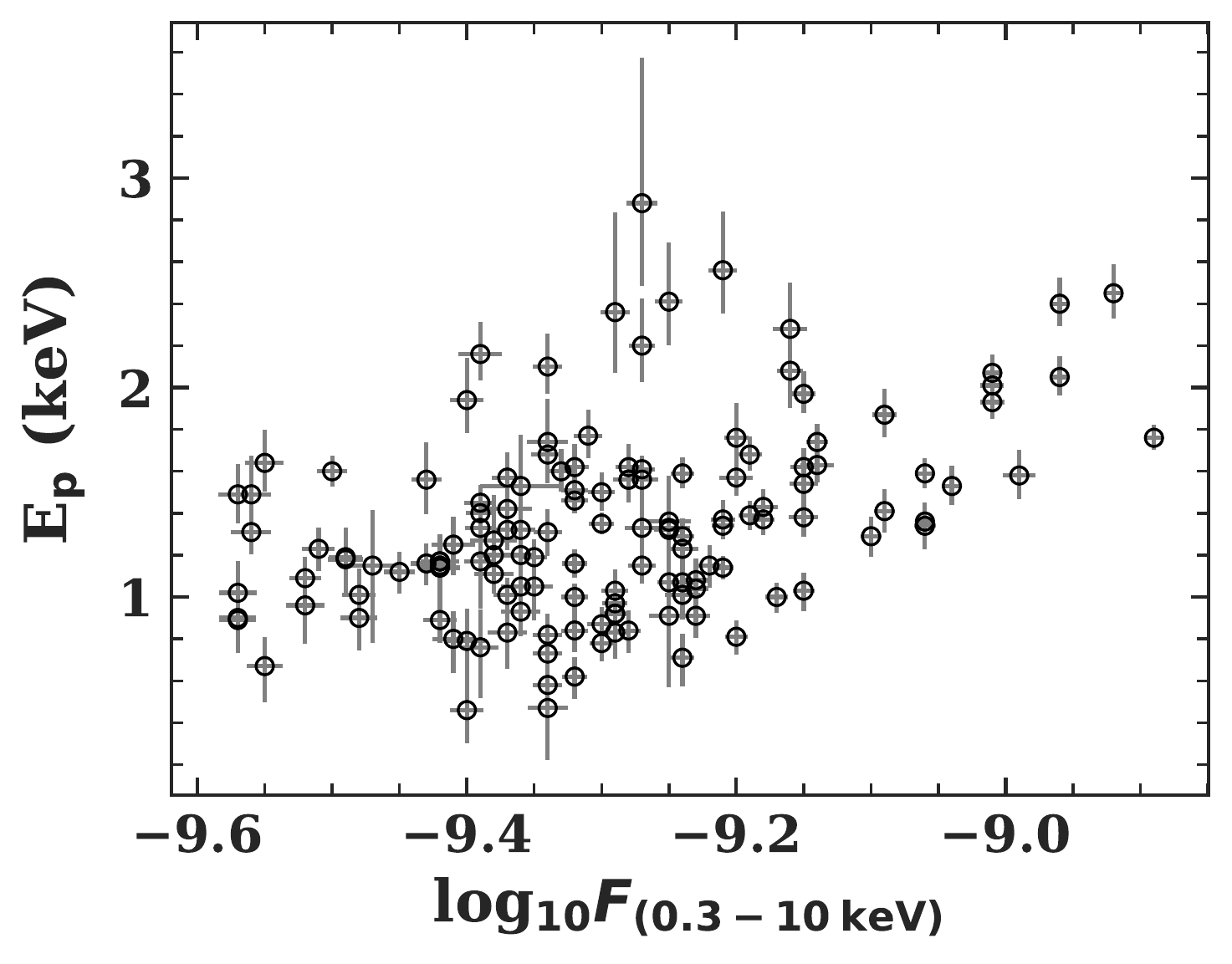}
\includegraphics[width=5.3cm , angle=0]{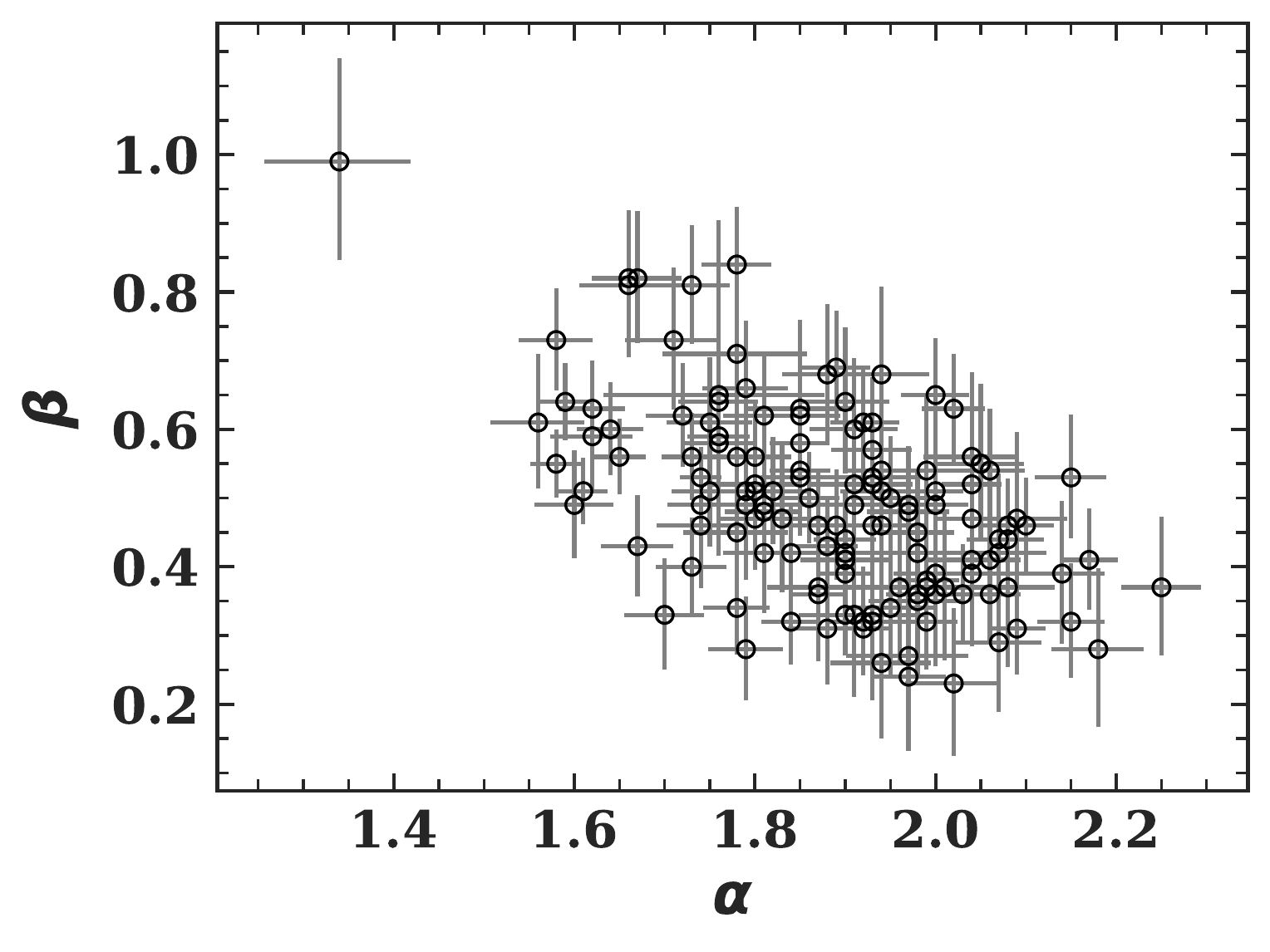}
\includegraphics[width=5cm , angle=0]{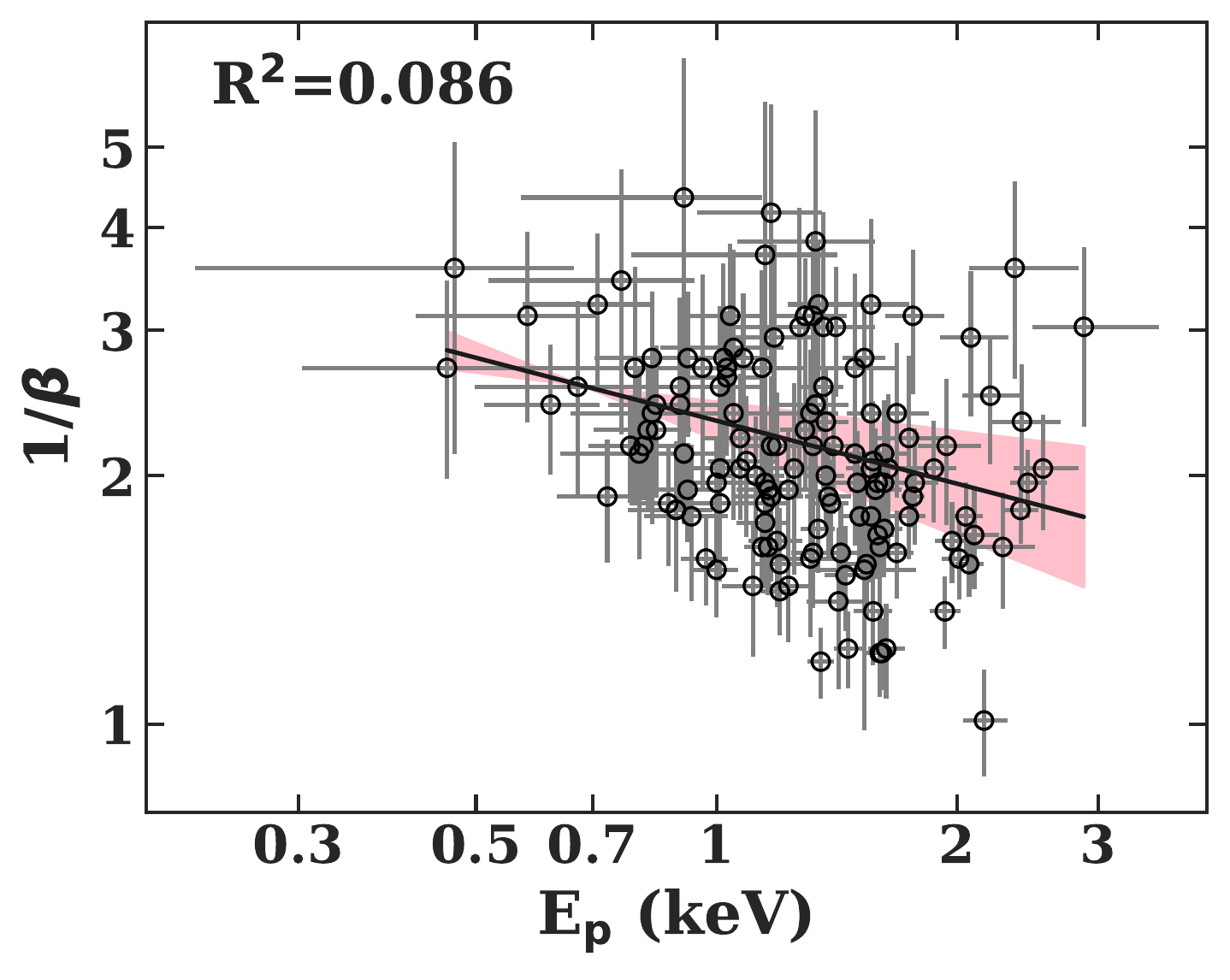}
\includegraphics[width=5cm , angle=0]{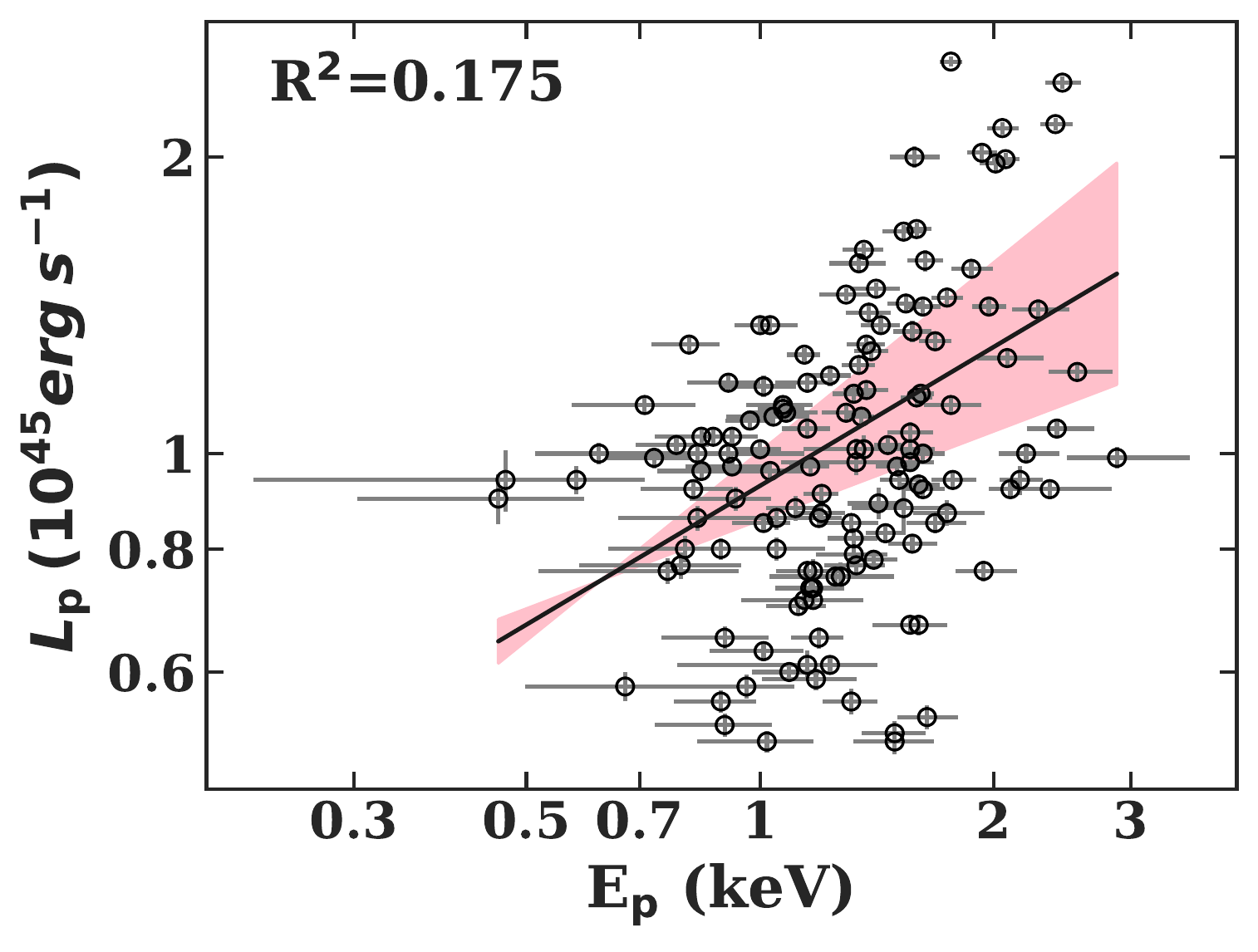}

\caption{Correlation between various spectral parameters of blazar 1ES 1959$+$650.}
\label{fig:parameter_plots}
\end{figure*}

\begin{table}
\centering
\caption {Results of $\chi^{2}$ test for Hardness Ratio analysis.}
\label{tab:chi}
\begin{tabular}{lcccc} \hline\hline
{Obs. ID}&{Obs. Date}& {DoF}& \boldmath{$\chi^{2}$}&\boldmath{$\chi^{2}_{0.90}$}\\
\hline
&&\boldmath{XMM-Newton}&&\\\hline
0850980101&2019-07-05&81&53.8&97.7\\
0870210101&2020-07-16&62&37.8&76.6\\\hline
&&\boldmath{Swift-XRT}&&\\\hline
00094153010 & 2018-07-02 & 16 & 3.4 & 23.5 \\
00095332009 & 2019-06-17 & 33 & 12.4 & 43.7 \\
00034588209 & 2020-06-21 & 21 & 7.0 & 29.6  \\
00013906004 & 2020-12-10 & 30 & 5.1 & 40.2 \\\hline\hline
\end{tabular}
\#Complete table of all observations appear in online supplementary material
\end{table}
 
\begin{figure*}
\centering
\includegraphics[width=7.0cm , angle=0]{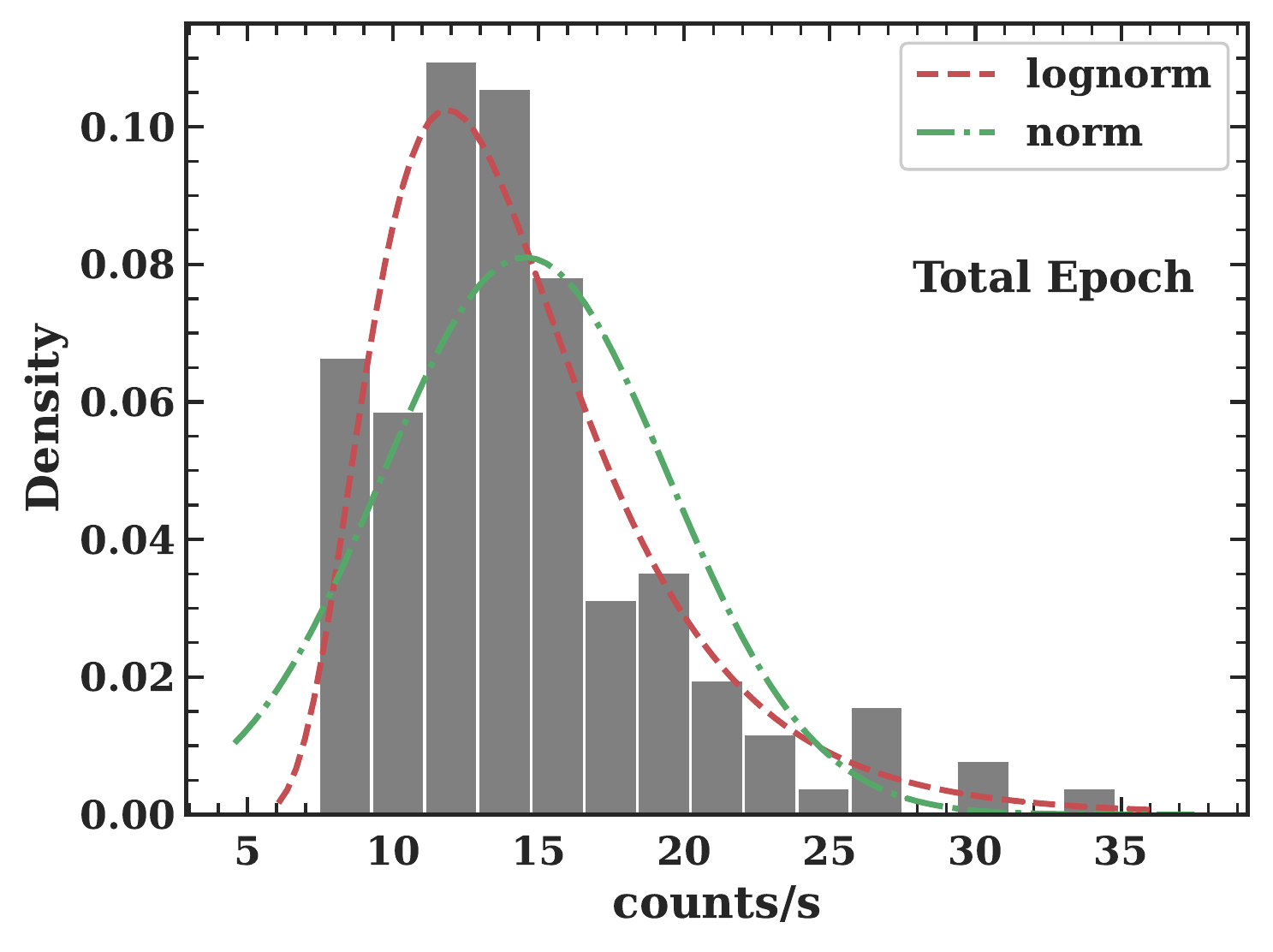}
\caption{Fitting of Log-normal (red dashed line) and Normal (green dash-dotted line) distributions to count rate histogram of total epoch.}\label{fig:Lognormality}
\end{figure*}
 
\section{Results}
\label{sec:4}
We studied 125 archived observations of the Swift satellite of the TeV blazar 1ES 1959+650 during the period June 2018$-$December 2020.
We also analyzed two publicly archived XMM--Newton observations of this blazar which are observed on 5th July 2019 and 16th July 2020.
These observations are used to study flux and spectral variability of this blazar on intra-day as well as on long term timescales. 

\subsection{Intraday flux and spectral Variability}
Blazar 1ES 1959+650 is studied with the XMM-Newton observations held on 5th July 2019 $\&$ 16th July 2020. The obtained light 
curves are shown in figure \ref{fig:xmm}. 
Flux variability is calculated using excess variance and amplitude is calculated to be 1.95 and 3.12 $\%$ respectively. 
We also analyzed Swift-XRT observations of this source during the period June 2018––December 2020 in total 125 nights. 
The sample of light curves are shown in figure \ref{fig:swift}. We found significant flux variability (i.e F$_{var} >$ 3$\sigma$) 
in five of these observations with amplitude varying between 4.11--7.34\% which are presented in Table ~\ref{tab:fvar}.

In order to calculate the spectral variability of these observations, HR analysis is performed which are presented
in table ~\ref{tab:chi}. HR analysis yields no spectral variability for any individual light curve which can be seen from the plots of
HR versus counts/s (0.3-10 keV) shown in figure \ref{fig:swift}. \\
\\
\subsection{Variability Timescale}
Doubling/halving timescale $\tau_{var}$ defined in equation \ref{eq:var_timescale} is used to calculate the shortest variability timescales. The shortest variability timescale we found as t$_{var}$ = 15.28 ks.
Emission region size is constrained with the equation defined as follows:
\begin{equation}
    R \leq \frac{\delta}{1+z} c \: {\tau_{\rm var}}
\end{equation}
with ${\tau_{\rm var}}$=15.27 ks, Doppler factor $\delta$=15 \citep{Patel_2018} and it is found to be 6.56 $\times$ 10$^{15}$ cm consistent with values found
in previous studies \citep{Magic2020, Shah_2021}.\\
\\
\subsection{Cross correlated variability and power spectral density}

Cross correlation studies are performed using XMM--Newton data as they are long observations with high cadence. The cross-correlation between the soft band  (0.3-2) keV and 
the hard band (2-10) keV is performed using DCF and their corresponding DCF plots are shown in the right panel of figure ~\ref{fig:xmm}. The DCF plots are fitted with a Gaussian function 
(as described 
in section \ref{sec:dcf}) and we have obtained a time lag of -0.94 and 0.36 ks respectively for the observations performed on 5th July 2019 and 16th July 2020 respectively. 
Results of DCF are provided in Table ~\ref{tab:dcf_psd}.

Significant correlation at positive/negative lags means that soft/hard variations are leading the variations in hard/soft bands 
respectively. During observation performed on 5th July 2019, the lags are negative i.e. the variations in the (2--10) keV are leading those 
in (0.3--2) keV. Therefore, in this case, variations at lower energies are slower than the variations at 
higher energies. The reverse situation is observed during the second observation. We found that there is a hard lag i.e. (2--10) keV
band is lagging behind the soft one (0.3--2) keV. X-ray emission of HSPs lie at the top of the synchrotron hump and are characterised
by the energy dependent acceleration and cooling mechanisms.

Following \cite{Zhang_2002}, the acceleration timescale $t_{\rm acc}$ and cooling timescale $t_{\rm cool}$  of the relativistic electrons in the observed frame 
can be expressed as a function of the observed photon energy $E$ (in keV):
\begin{equation}
t_{\rm acc}(E) = 9.65 \times 10^{-2} (1+z)^{3/2} \xi 
	B^{-3/2} \delta^{-3/2} E^{1/2} \quad {\rm s} \,,
\label{eq:tacc}
\end{equation}
\begin{equation} 
t_{\rm cool}(E) = 3.04 \times 10^{3} (1+z)^{1/2} B^{-3/2}
	\delta^{-1/2} E^{-1/2}  \quad {\rm s} \,,
\label{eq:tcool}
\end{equation}
where $z$ is the source's redshift, $B$ is the magnetic field in Gauss, $\delta$ is the Doppler factor of the emitting region, and $\xi$ is the parameter describing how fast the electrons can be accelerated.
 
One can note that $t_{\rm acc}$ and $t_{\rm cool}$ both depend on the photon energy in inverse fashion. The higher energy electrons 
cool faster as compared to lower energy electrons but accelerated slower than lower energy electrons. Therefore, if  $t_{\rm cool}$ is
 greater than $t_{\rm acc}$, cooling process dominates \citep{Kirk_1998}. In such a case, higher energy photons will lead to lower energy photons and soft lag is expected.

\begin{equation}
\tau_{\rm soft} = t_{\rm cool}(E_{\rm l}) - t_{\rm cool}(E_{\rm h}),
\label{eq:softlag}
\end{equation}

If $t_{\rm acc}$ is comparable to $t_{\rm cool}$ in the observed energy range, acceleration processes dominates in the emitting region and hard
lag is expected. The time lag in an acceleration dominated system is expressed as:

\begin{equation}
\tau_{\rm hard} = t_{\rm acc}(E_{\rm h}) - t_{\rm acc}(E_{\rm l}),
\label{eq:hardlag}
\end{equation}

Using the observed lags, one can discern the physical parameters of the emitting region as follows:

\begin{equation}
B\delta \xi^{-2/3} = 0.21 \times (1+z) E_{\rm h}^{1/3} 
    \left [ \frac{1 - (E_{\rm l}/E_{\rm h})^{1/2}}
                 {\tau_{\rm hard}} \right ]^{2/3} {\rm Gauss},
\label{eq:hard}
\end{equation}
\begin{equation}
B\delta^{1/3} =209.91 \times \left (\frac{1+z}{E_{\rm l}}\right
)^{1/3}
	\left [\frac{1 - (E_{\rm l}/E_{\rm h})^{1/2}}
        {\tau_{\rm soft}} \right ]^{2/3}{\rm Gauss}.
\label{eq:soft}
\end{equation} 
where $\tau_{\rm hard}$ and $\tau_{\rm soft}$ refer to the observed hard and soft lags (in second) between the low $E_{\rm l}$ and 
high $E_{\rm h}$ energy bands (in keV), respectively, ($E_{\rm l}$ and $E_{\rm h}$ are logarithmic averaged energies of the given energy 
bands).

Equation \ref{eq:soft} is used to calculate the
magnetic field with redshift z=0.048, ${\tau_{\rm soft}}$=940 sec and Doppler factor $\delta$=15 \citep{Patel_2018}. Magnetic field, B of the emitting region is found to be 0.64$\pm$0.05 Gauss which is found to be consistent
with the values provided in the literature.
\cite{Tagliaferri_2003} and \cite{Magic2020} performed SED modeling and obtained magnetic field
close to the value we have obtained.

\subsection{Long Term flux and spectral Variability}
We study long term flux and spectral variability of blazar 1ES 1959+650, observed between the period June 2018$-$December 2020. Depending on the flux state of the blazar, 
we have divided the total epoch into three periods. 
The flux state of the source is defined with respect to the average counts during our observing run.
The source is in high state during period 1 [MJD 58301.75--58577.86] which is during 2018 July$-$2019 April 04), in an intermediate state during period 2 [MJD 58581.37--58920.97]
 which is 2019 April 08$-$2020 March and in low flux state during period 3 [MJD 59001.42--59208.88] which is June$-$December 2020 with the mean count rate of 18.28, 14.82 and 10.90 counts/s respectively. The long term light curve of the blazar 1ES 1959+650, observed during the period June 
2018$-$December 2020, is shown in figure ~\ref{fig:longterm_lc}. The light curves for different periods are shown in figure 
~\ref{fig:temporal_var}. During long term timescales, the source showed significant spectral variability. The HR, 
$\alpha$, $\beta$ and E$_{p}$ have showed long term variations which can be seen in figure ~\ref{fig:temporal_var} 
for different periods.

In period 1, source exhibit two flares with flux reaches upto $128.82 \times 10^{-11}\:{\rm erg\:cm^{-2}\:s^{-1}}$ on 2018-09-16. Fractional
variability amplitude $F_{var}$ is found to be $\sim$29.96\% and significant spectral variability is also found during this period. Photon index has hardened with increase 
in flux of the source which varies between 2.17--1.58. Lower values of curvature parameter 
are found which ranges between
0.84--0.31. Peak energy shows positive correlation with respect to flux of the source and reached upto 2.45 keV. 
During Period 2, source exhibits small flares with mean flux level of this period reaching upto $52.93 \times 10^{-11}\: {\rm erg\:cm^{-2}\:s^{-1}}$. 
$F_{var}$ is found to be 23.12\% and significant spectral variability is also found during this period which can be seen in Fig ~\ref{fig:temporal_var}. 
Photon index has hardened with increase in flux of the source which varies between 2.25--1.56. Similar to Period 1, $\beta$ 
has lower values 
which ranges between 0.82--0.23. During Period 3, source exhibit relatively low flux state with highest flux reaching upto $61.66 \times 10^{-11}\: {\rm erg\:cm^{-2}\:s^{-1}}$.
$F_{var}$ is found to be 21.43\% and significant spectral variability is found during this period.
Photon index has hardened with increase in flux of the source which varies between 2.14--1.34. $\beta$ has lower values ranging
between 0.99--0.24. $E_{p}$ is positively correlated with flux and reaches upto 2.88 keV. \\
\\
\\
\subsection{Log Normality of flux distributions}
The nature of variability in a long term period can be quantified or explained with the flux distribution of the variable 
source \citep{Uttley2005}. Blazars exhibits a log-normal flux distribution over the normal distribution \citep{Uttley2005, Chakraborty2020, shah2020} which could be an indication 
of the variability imprint of the accretion disk onto the jet. The flux distribution is expected to be a Gaussian for a linear stochastic process, while it is expected to be a lognormal
 for multiplicative processes that originate in the accretion disk \citep{Uttley2005}.
The fluctuations of the log-normal fluxes are proportional to the flux itself and indicates underlying multiplicative physical processes.
The lognormality of blazars on different time scales and in different spectral ranges are studied many times in literature \cite[i.e.][]{Kushwaha_2016, Sinha_2017, Chevalier_2019, 
Bhatta_Dhital_2020}. However, the dominance of doppler boosted jet emission in blazars restricts our understanding of the accretion disc-jet connection.
 \cite{Kushwaha_2016} performed an extensive study of lognormality of flux distribution of blazars for the first time. Specifically, variations on minutes/hours like timescales 
should be independent of accretion disc perturbations and favours originating from the instabilities in the relativistic jets \citep{Gaidos_1996, Albert_2007, Narayan_Piran_2012}. 
As blazars have strong magnetized jets pointing towards us, flux distributions using minijet-in-a-jet model \citep{Giannios_2009} are studied 
by \cite{Biteau_Giebels_2012} and they found that the flux from a single randomly oriented mini jet will follow a Pareto distribution. The flux integrated from many isotropically
 oriented mini jets could lead to an $\alpha$ stable distribution which could converge to a log normal distribution when subjected to experimental uncertainties. 
Therefore, non-linear flux distributions can arise from small Gaussian perturbations. This could provide an explanation for the lognormal flux distributions in blazars
during flaring states. In this scenario, flux distribution has been found to hold the rms-flux relation \citep{Biteau_Giebels_2012}.

An alternative interpretation for the non-Gaussian distribution of blazar variability light curves is provided by \cite{Sinha_2018}. They explained it using a small perturbation 
in the acceleration timescale which can result in the variability of the particle number density that is a linear combination of Gaussian and lognormal processes. 
The dominant shape of the resultant flux distribution is determined by relative weight of these processes \citep{Sinha_2018, Bhatta_Dhital_2020}. 
They also demonstrated that perturbation in the acceleration time-scale leads to Gaussian distribution in its photon index, whereas perturbation in the particle cooling rate 
produces neither of these distributions \citep[][and references therein]{Sinha_2018, shah2018, shah2020, Khatoon_2020, Khatoon_2022}.

Flux distribution of 1ES 1959+650 is studied by \cite{Patel_2018} using radio to $\gamma$-ray data. It is also studied by \cite{Bhatta_Dhital_2020} using decade-long Fermi/LAT 
observations. \cite{Duda_Bhatta_2021} used maximum likelihood estimation (MLE) methods to study flux distributions.

In order to investigate log normality of flux distribution in our observations, we fit the histograms of the X-ray data observed by Swift-XRT 
between the period June 2018 to December 2020 with the Gaussian and log-normal distributions. The figure~\ref{fig:Lognormality} shows the log-normal and normal 
flux distribution of this blazar for the total epoch used in this analysis.
We have used Anderson-Darling (AD) test \cite[e.g.][]{AD_1952, AD_1954, Jantschi_2018, Stephens_1977, DAgostino_Stephens_1986, shah2018} to quantify the nature of 
flux distribution of blazar 1ES 1959+650. We have obtained that the blazar 1ES 1959+650 follows a log-normal behaviour in total epoch with a 
$p$-value = 0.20. We also found a significant linear correlation between rms and flux for the total epoch which indicates that the variability might arise from the minijet-in-a-jet model. However, as recently shown by \cite{Scargle_2020}, the linear rms-flux relationship can be obtained from intrinsically additive processes, therefore this result might 
be used with caution.
We also calculated the photon index distribution for the total epoch and found it to be well fitted with lognormal as well as normal distributions. As discussed in \cite{Sinha_2018}, 
small temporal fluctuations in the intrinsic time-scales in the acceleration region is capable of producing particle distributions with non-Gaussian signatures and 
significant flux-rms correlations. Therefore, we cannot rule out the possibility of an acceleration-due-to-shock scenario.

\subsection{Relation between spectral parameters}
\label{sec:4.6}

Blazar 1ES 1959+650 spectra are fit by log parabolic and power law models. The results of both the models are
presented in table \ref{tab:spectral_fitting}. We used the F-test to compare the fitting results of these two models. We found that all the spectra are 
well fitted by a log parabolic model.
Then, we derived the spectral fitting parameters i.e. location of synchrotron peak($E_{p}$), peak luminosity ($L_{p}$) with the log parabolic model. The results show the variation in $E_{p}$ in
the range of 0.46$-$2.88 keV whereas $L_{p}$ varies in the range of 0.51$-$2.50 ${\rm erg\:cm^{-2}\:s^{-1}}$. 
Spectral variations of the photon index varies between 1.34$-$2.25 and the curvature ($\beta$) varies between 0.21$-$0.99.
Correlation between various spectral parameters are shown in figure \ref{fig:parameter_plots} and are studied with 
Spearman's rank correlation coefficient $\rho_{s}$ and their corresponding $p$-values. The
results are presented in table \ref{tab:spearman_coeff}.

A positive correlation is expected between photon index $\alpha$ and curvature $\beta$ in the first order fermi acceleration scenario. This correlation
is predicted for the energy dependent acceleration probability process EDAP (i.e. \citealt{Massaro(2004)}) where the probability
$p_{i}$ that a particle undergoes an acceleration step $i$, with the corresponding energy $\gamma^{q}_i$
and energy gain $\varepsilon$, is given by $p_i$=$g/\gamma^{q}_i$ where $g$ and $q$ are positive constants.
Therefore, as the energy of the particle increases, the probability of the particle's acceleration decreases. According to \cite{Massaro(2004)},
a linear relationship is expected between spectral index \emph{s} and curvature \emph{r}, $s=-r(2/q)\log{g/\gamma_0}-(q-2)/2$. 
The synchrotron emission is produced by the differential energy spectrum of the form 
$N(\gamma)\sim{\gamma/\gamma_0}^{-s-r\log{\gamma/\gamma_0}}$ is given by 
$P_S (\nu)\varpropto (\nu/\nu_0)^{-(a+b\log(\nu/\nu_0))}$ with $a=(s-1)/2$ and $b=r/4$ (i.e. \citealt{Massaro(2004)}).
In our observations, we found weak negative correlation between $\alpha$ and $\beta$ which is expected when
$g>\gamma_0$ (i.e. \citealt{Kapanadze_2020}). It implies that there exists electron population with a very low initial energy 
$\gamma_0$ in the emission zone. \cite{Kapanadze_2020} found a negative correlation between these quantities for Mrk 421 for some
of its observational period during 2015 December--2018 April. 
The co-existence of second 
order Fermi acceleration/stochastic acceleration could also weaken the $\alpha$--$\beta$ correlation. \cite{Katarzynski_2006}
have shown via simulations that electrons can be accelerated at the shock front via EDAP but can gain energy via the stochastic 
mechanism after escaping the shock front. The combined effect of both such processes could result in a weak or no strong
correlation between $\alpha$ and $\beta$. \citep{Kapanadze2016, Kapanadze_2018c, Kapanadze_2020}.

The synchrotron peak energy ($\textit{E}_\textrm{p}$) and the luminosity ($\textit{L}_\textrm{p}$) of a source follows
 a power-law relation of the form $\textit{L}_\textrm{p} \propto \textit{E}_\textrm{p}^{\:a}$ (i.e. \citealt{Rybicki_Lightman_1979}). 
 If the electron distribution in the emitting region follows a log-parabolic distribution, the peak luminosity is given by 
$\textit{L}_\textrm{p} \sim \textit{N} \gamma_{\textrm{p}}^\textrm{2} \textit{B}^\textrm{2} \delta^\textrm{4}$.
The peak energy follows $\textit{E}_\textrm{p} \sim \gamma_{\textrm{p}}^\textrm{2} \textit{B} \delta$ (e.g., \citealt{Tramacere_2009}). $\gamma_{\textrm{p}}$ represents the peak of $\textit{n}\textrm{(}\gamma\textrm{)} \gamma$ where $\gamma$
is the electron Lorentz factor; 
$\textit{N} \sim \emph{n}(\gamma_\textrm{p})\gamma_\textrm{p}$ is the total electron number, \textit{B} represents the magnetic field and 
$\delta$ is the Doppler beaming factor. If $a$ = 1, the spectral changes are mainly caused by the variations of 
the average electron energy, but the total electron number remains constant; if $a$ = 1.5, they are mainly 
caused by the variations of the average electron energy, but the total electron number also changes; if $a$ = 2, they are correlated with the changes of magnetic field, $B$; 
and if $a$ = 4, they might be dominated by the variations of the beaming factor, $\delta$.
We found significant correlations between $\textit{E}_\textrm{p}$ and $\textit{L}_\textrm{p}$ during our observations.
We fit our observational data using the equation $\log {\textit{L}_\textrm{p}}=a \log {\textit{E}_\textrm{p}}+b$
and found $a$ to be $<$1 during the whole observational period. During period 1, $a$ $\sim$ 1 which concludes that the 
spectral changes might be caused by the variations of the average electron energy.

The correlation between $\textit{E}_\textrm{p}$ and $\beta$ provides us 
clues about the acceleration mechanism  \cite[e.g.][]{Massaro(2004), Tramacere_2011} whether it is statistical/stochastic mechanisms. These two mechanisms are produced by the log-parabolic electron distribution, resulting in a log-parabolic
SED.

In statistical acceleration process, the electron energy distribution follows the log-parabolic law and the acceleration 
efficiency of the emitting electrons is inversely proportional to their energy \citep{Massaro(2004)}. Therefore, in such process, 
$\textit{E}_\textrm{p}$ and \textit{$\beta$} follow the correlation of the form $\log \textit{E}_\textrm{p} \approx \textit{Const.}+\textrm{2}/
\textrm{(5}\textit{$\beta$}\textrm{)}$, with the assumption of \textit{$\beta$} = \textit{r}/4 where $r$ is the curvature of the electron 
energy distribution \citep{Chen_2014}. For the fluctuations of fractional acceleration gain process, 
electron energies follow log-normal law, and the energy gain fluctuations are a random variable around the 
systematic energy gain \citep{Tramacere_2011}. In such case, $\textit{E}_\textrm{p}$ and \textit{$\beta$} follow the correlation 
of $\log \textit{E}_\textrm{p} \approx \textit{Const.}+\textrm{3}/\textrm{(10}\textit{$\beta$}\textrm{)}$ given \textit{$\beta$} = 
\textit{r}/4 \citep{Chen_2014}.

Second scenario is described by the stochastic acceleration process. Here, a momentum-diffusion term is included in the kinetic equation, 
which leads to energy gain fluctuations in the diffusive shock acceleration process \citep{Tramacere_2011}. 
In this process, $\textit{E}_\textrm{p}$ and \textit{$\beta$} follows the relation of $\log \textit{E}_\textrm{p} \approx \textit{Const.}
+\textrm{1}/\textrm{(2}\textit{$\beta$}\textrm{)}$ where \textit{$\beta$} = \textit{r}/4 \citep{Chen_2014}.
Therefore, theoretically expected values of $C$ are 10$/$3, 5$/$2 and 2 for the fractional acceleration gain fluctuation, 
energy-dependent acceleration probability and stochastic acceleration processes, respectively.
We found a significant negative correlation between $\textit{E}_\textrm{p}$ and \textit{$1/\beta$} in our observations.
We fit our observational data using the equation $\textrm{1}/\textit{$\beta$}=\textit{C} \log {\textit{E}_\textrm{p}}+\textit{D}$
but did not get the value of $\textit{C}$ close to the above values hence we cannot explain our observational data using any of the
above acceleration mechanism. However, the co-existence of the above acceleration mechanism might be possible \citep{Wang_2019} which could lead to overall weakening
of correlation.

A positive correlation between flux and $\textit{E}_\textrm{p}$ is found during our observations which indicates the shift of synchrotron peak
to higher energies. Also, we found significant correlation between flux and $\textit{S}_\textrm{p}$ which is expected 
as peak height increases as flux increases \citep{Holder_2003, Kapanadze2016, Kapanadze_2016b, Kapanadze2018a, Kapanadze2018b, Wang_2019, Chandra_2021}.
Near the peak energy of the emission, the cooling timescale shortens and can complete with the acceleration timescales \citep{Tramacere_2009} 
which leads to an anti-correlation between $\textit{E}_\textrm{p}$ and \textit{$\beta$} if the cooling timescales is shorter than that of 
EDAP or stochastic acceleration.

We found significant correlation between flux and $\alpha$ which is an indication of hardening of the spectra as flux increases
which is very common for HSP type blazars. 
It indicates that the  hard X-rays are varying more rapidly than soft X-rays \citep{Cui_2004, Giebels_2002, Tagliaferri_2003, 
Holder_2003} or there is an injection of fresh electrons with an energy distribution harder than the
previously cooled electrons \citep{Mastichiadis_2002}.

\section{Discussion}
\label{sec:5}
We observed the blazar 1ES 1959$+$650 during the period June 2018--December 2020 when the source showed different flux states (high/low)
to study their flux and spectral variability on intra-day and long term timescales using SWIFT satellite. The source is studied for
intra-day flux variability in total 125 nights and found significant variability in only five nights. We did not find any spectral variability during
 these observations. The source is studied using two observations of XMM--Newton held on 5th July 2019 and 16th July 2020 and
found significant flux variability in both of these observations with low amplitude variation of 1.95$\%$ and 3.12$\%$ respectively. 
The source is favoured by the shock-in-jet model where IDV is triggered by the interaction of the shock front with jet inhomogeneities;
turbulence behind a shock front; smallest-size jet turbulent structures (i.e. \citealt{Marscher_Gear_1985, Wagner_1995, Sokolov_2004}). Smallest size jet structures are attributed to produce very rapidly variable emission due to light
travel arguments. We found flux doubling timescale of 15.27 ks (between MJD 58670.07--58670.13) which leads to size of the emitting region to be 6.56 $\times$
10$^{15}$ cm and black hole mass estimate of 2.95 $\times$ 10$^{8}$ M$_\odot$ which is obtained as follows:
\begin{equation}
    M_{BH}=\frac{c^{3}t_{var}}{10G(1+z)}
\end{equation}
where G is the Gravitational constant \cite[e.g.][]{Zhang_2021}). \cite{Yuan_2015} reported the optical timescales in the range
from 23 minutes to 3.72 hr with the Kerr
black hole mass in the range (0.42--4.09) $\times$ 10$^{8}$ M$_\odot$. \cite{Falomo_2003} used the host
galaxy luminosity relation and \cite{Kurtanidze_2009} used the variability timescales to estimate the black hole mass to be 3.16 $\times$ 10$^{8}$ M$_\odot$.

XMM--Newton observation of 1ES 1959$+$650 during two observations suggest the presence of soft as well as hard lags which infers
that $t_{\rm acc}$ and $t_{\rm cool}$ of the emitting region changes from epoch to epoch. This is consistent with previous studies.
Hard lags suggest that the flux evolution is dominated by acceleration processes while soft lags suggest that the emission region is
dominated  by cooling mechanisms. The first-order Fermi acceleration process \citep[e.g. ][and references therein]{Kirk_1998} and statistical/stochastic processes 
involving second-order Fermi acceleration \citep{Katarzynski_2006, Becker_2006} are most acceptable models
for particle acceleration and therefore, hard delays are modulated by variations of the acceleration parameter $\xi$. We have estimated the value of magnetic field to 
be 0.64$\pm$0.05 Gauss using the soft lags and the values are found to be close to those reported in literature for our source.

PSD analysis is done using two observations of XMM--Newton which are used to characterize the variability on intra-day timescales.
Accretion disc models typically produce PSD slopes in the range between \mbox{-1.30 -- -2.10} \citep{Zhang_Bao_1991, Mangalam_Wiita_1993, Kelly_2011} while jet based models yield steeper slopes in 
the range between \mbox{-1.70 -- -2.90} (i.e. \citealt{Calafut_wiita_2015, Pollack_2016, Wehrle_2019}). The observed PSD slopes 
of -2.41 and -2.15 for intra-day light curves are steeper as compared to those predicted by accretion disk models and are more consistent with jet based 
models. However, the small number of PSDs used here provides a tentative hint favouring fluctuations originating in jets. 

 On long term timescales, source exhibit high state during Period 1 with two flares with flux reaching upto 34.82 counts/s.
During Period 2, the flux of the source decreases to 10.35 counts/s and during Period 3, the flux of the source is lowest
with reaching upto 7.43 counts/s. Source showed significant spectral variability throughout our observations. 
We studied flux distributions of our source during different observational periods and found that the source
follows a log-normal behaviour during the total epoch. 
\cite{Kapanadze_2020} studied the lognormality of Mrk 421 using the multi-wavelength data and found that log-normal fits were preferred over normal fits for most of their dataset.
The flux variability of the source is attributed to the propagation of shocks downstream the relativistic jets \citep{Sokolov_2004}. The formation of these shocks might be 
related to the turbulence/inhomogeneities occuring in the accretion disk \citep{Kushwaha_2016, Sinha_2017, Kapanadze_2020NewA_b, Kushwaha_2020}. However, this is not always the case.
Many observations including flaring events showed deviations from log-normal distribution which indicates that these flares might be triggered by the interaction of shocks
with the jet inhomogeneities which could be related to the jet instabilities (i.e. \citealt{Marscher_2014}).   

All the spectra are well described by log parabolic model yielding spectral curvature ranging between 0.23--0.99 and photon 
index varies between 1.34--2.25. During our observations, peak energy $E_{p}$ varies in such a way 
that $E_{p}$ shifts upto higher energies as flux increases. We studied correlation between the spectral parameters derived from
log parabolic model. We found weak negative correlation between $\alpha$ and $\beta$ which is due to the co-existence of 
stochastic and statistical acceleration processes \citep{Kapanadze2018b, Kapanadze_2018c}. 
The spectral hysteresis analysis of 1ES 1959+650 showed an interplay between the acceleration and cooling timescales
of emitting particles and ﬂux variability timescale \citep{Kapanadze2018a, Kapanadze2018b}.
Correlation between $E_{p}$ and $L_{p}$ also
follows a powerlaw as described in section \ref{sec:4.6}. During Period 1, we found spectral changes to be caused by the variations 
of the average electron energy. The anti-correlation between $E_{p}$ and $\beta$ is expected for the efficient
stochastic acceleration of electrons by the magnetic turbulence which is not seen in our observations. The weak correlation between 
these parameters implies the co-existence of stochastic and statistical acceleration processes in the emitting region. \\
\\
\section{Conclusion}
 \label{sec:6}
The main findings of this work are summarized as follows:
\begin{enumerate}
\item[{\bf 1.}] Swift-XRT and XMM--Newton EPIC-pn observations have been used to study the HSP 1ES 1959$+$650  during the period 
June 2018--December 2020 in total 127 nights of observations. Significant variability is detected in total 7 of the nights with 
flux variability amplitude varying between 1.95--3.12 \%. Hardness ratio analysis shows no significant spectral variability in any of the nights.
The flux doubling timescale is found to be 15.27 ks and the black hole mass is calculated to be 2.95 $\times$ 10$^{8}$ M$_\odot$.

\item[{\bf 2.}] Using XMM--Newton observations, cross-correlation between soft band (0.3--2) keV and hard band (2--10) keV 
were performed using the DCF method.  Both DCF plots are correlated and hard as well as soft lag of ~360 and 940 seconds respectively 
are found which provides magnetic field strength of $\sim$ 0.64$\pm$0.05 Gauss in the jet.

\item[{\bf 3.}] PSD analysis is performed using XMM--Newton observations and power law slopes are found to be -2.41 and -2.15
which favours jet based model (as discussed in section \ref{sec:5}).

\item[{\bf 4.}] Intra-day light curves are checked for lognormality behaviour and found that they are well modelled by normal as well as lognormal distributions. As suggested by 
\cite{Gaidos_1996, Narayan_Piran_2012}, variations on minutes/hours like timescales are independent of accretion disc fluctuations and could be attributed to some linear/non-linear perturbations in the physical parameters used to model the relativistic jets in blazars. The most plausible model to explain short-term variability is turbulence in the jet behind the reconfinement
shock that contains multiple synchrotron emitting cells of different sizes within the single emitting region which yields impressive light curves, PSDs and polarization variations (i.e. \citealt{Pollack_2016, Marscher_2014}).

\item[{\bf 5.}] Source exhibits log normality behaviour on long term timescales. Long term variations are explained by superposition of small flares on long term trends.
As blazar jets are highly magnetized, variability may be incorporated by minijets-in-a-jet model as we found a linear
correlation between rms--flux relation. However, as the photon index distribution is well fitted by Gaussian as well as normal distributions, we can not rule out the possibility of 
propagation and evolution of relativistic shocks through the jet leading to variability in X-ray bands. These shocks could be related to an abrupt increase of the plasma injection rate at
the jet base owing to the instabilities in the accretion disc.

\item[{\bf 6.}] On the long timescales, source showed high as well as low flux states. Log parabolic model is required to describe
the X-ray spectra of this source yielding spectral curvature values ranges between 0.23--0.99 and photon index ranges between 1.34–2.25.
Position of synchrotron SED peak $E_{p}$ varies between 0.46–2.88 keV. Synchrotron peak $E_{p}$ strongly correlates with flux which
 implies that $E_{p}$ increases as flux increases. Hardness ratio analysis on long term timescales indicates that the source
follows the 'harder-when-brighter' trend.

\item[{\bf 7.}] Source showed weak correlation between photon index $\alpha$ and curvature $\beta$ which could be due to the combined effect of statistical/stochastic processes.
Electrons can be accelerated at the shock front via EDAP but they can gain energy via the stochastic mechanism after escaping the shock front \citep{Katarzynski_2006}.

\item[{\bf 8.}] 1ES 1959$+$650 showed a low spectral curvature ($\beta$ $\sim$ 0.23--0.99) and an anti-correlation between $E_{p}$ 
versus 1/$\beta$ is found which might be due to the co-existence of stochastic and statistical acceleration processes.

\end{enumerate}

\section*{Acknowledgements}
We would like to thank the anonymous reviewer for the constructive comments that helped us to improve the paper scientifically.
 We acknowledge the use of public data from the Swift data archive. This research is based on observations obtained with XMM-Newton, an ESA science mission with instruments and contributions directly funded by ESA Member States and NASA.
KAW and HG acknowledge the financial support from the Department of Science and Technology, India through INSPIRE  Faculty
award IFA17-PH197 at ARIES, Nainital.

\appendix
\label{appendix:a}
For an evenly sampled light curve (with a sampling period $\Delta$T) comprising a series of fluxes $x_{i}$ measured at discrete times $t_{i}$ (i = 1, 2, . . . , N ):
\begin{equation}
|DFT(f_{j})|^2= \bigg|\sum_i x_{i}\: e^{2\pi i\:f_{j}\:t_{i}}\bigg|^2 \:,
\end{equation}

at N/2 evenly spaced frequencies $f_{j} = \frac{j}{N\: \Delta T}$ (where j = 1, 2, . . . , N/2), $f_{N/2} = \frac{1}{2\: \Delta T} $ is the Nyquist frequency, $f_{Nyq} $ which denotes the maximal frequency that can be meaningfully inferred.

To eliminate the zero frequency power it is important to subtract the mean flux from the light curve before calculating the DFT \citep{Vaughan(2003b)}. 

PSD is then defined as \citep{Vaughan(2005)}:
\begin{equation}
P(f_{j})=\frac{2\Delta T}{N \bar{x}^{2}}\:|DFT(f_{j})|^2\:.
\end{equation}

The log-parabolic model is given by

\begin{equation}
F(E) = K(E/E_{1} )^{(-\alpha-\beta log(E/E_{1} ))}\:,
\end{equation}

in units of photons cm$^{-2}$ s$^{-1}$ keV$^{-1}$ (e.g., \citealt{Massaro(2004)}). $E_{1}$ is the reference energy, generally fixed to 1 keV. The parameter $\alpha$ is the spectral index 
at the energy of $E_{1}$ , while $\beta$ is the curvature parameter around the peak. $K$ is the normalization constant. The location of the synchrotron peak is calculated by
\begin{equation}
E_{p,logpar} = E_{1} 10^{(2-\alpha)/2\beta} \text{(keV)}\:,
\end{equation}

Another form of log-parabolic model i.e. {\it eplogpar} model is used to calculate synchrotron peak E$_{p}$. It is defined as
\begin{equation}
F(E) = K10^{-\beta(log(E/E_{p} ))^{2}}/E^{2}\:,
\end{equation}
in units of photons cm$^{-2}$ s$^{-1}$ keV$^{-1}$. (e.g. \citealt{Tramacere_2007,Tramacere_2009}).
E$_{p}$ is the synchrotron peak in units of keV, $\beta$ is the curvature parameter, which is the same as the parameter $\beta$ in the above log-parabolic model. The parameter 
K is the flux in $\nu$F$_{\nu}$ units at energy E$_{p}$ keV.

\bibliographystyle{aasjournal}
\bibliography{new_ms}{}
\end{document}